\documentclass[%
 reprint,
 amsmath,amssymb,mathrsfs,
 aps,
]{revtex4-2}
\usepackage[utf8]{inputenc}
\usepackage{physics}
\usepackage{mathrsfs}
\usepackage{graphicx}
\usepackage{mathtools}
\usepackage{caption}
\usepackage{bibentry}
\usepackage{hyperref}
\usepackage{braket}
\usepackage{amssymb}
\usepackage[title]{appendix}

\newcommand{\uev}{\ensuremath{\mu \rm{eV c^{-2}}}}

\begin{document}

% \preprint{APS/123-QED}

% \title{Signatures of Non-Classical Features of the Axion Field in Cavity Haloscopes}
\title{Signatures from Non-classical Features of Axion Dark Matter in Cavity Haloscopes}

\author{Erik W. Lentz}
\email{erik.lentz@pnnl.gov}
\affiliation{Pacific Northwest National Laboratory, 
            902 Battelle Blvd., 
            Richland, WA 99352, USA}

\date{\today}

\begin{abstract}

I present in this article how non-classical features in the local axion dark matter state may imprint on cavity haloscopes. Example signatures from squeezed and pair-entangled DM states are presented and their potential for sensitivity enhancements of experimental searches attuned to these signatures compared to traditional classical field signatures are discussed.

\end{abstract}


\maketitle{}

%\tableofcontents

\section{Introduction}
\label{sec:introduction}

The existence of abundant dark matter (DM) and its nature are unanswered questions, highlighting fundamental gaps in our physical understanding of the Universe~\cite{P5report2023}. The (invisible) axion presents a compelling DM candidate. Furthermore, it also resolves the unobserved-yet-expected charge-parity violation from the strong force theory QCD, commonly known as the strong CP problem~\cite{PQ1977,Weinberg1978,Wilczek1978,ABBOTT1983,PRESKILL1983,DINE1983}. This so-called QCD axion is an excellent candidate for DM due to its light mass and largely cold and abundant creation mechanism resulting in a highly degenerate Bose fluid, possibly a Bose-Einstein condensate (BEC), often referred to as ``wave-like'' or ``Bose'' DM \cite{Sikivie2009,Erken2012,Banik2016,Berges2015,Davidson2015,Guth2015,Marsh:2015xka,Levkov_2018,Eggemeier_2019,Eggemeier_2020,Mocz2018,Schive_2014,Veltmaat_2018,Veltmaat_2020,Lentz2019,Lentz2020,jaeckel2022reporttopicalgroupwave}. For axions to saturate the Concordance Model DM density~\cite{Planck2018}, numerical and analytical studies of QCD prefer an axion mass in the 1--100~\uev mass range~\cite{Berkowitz2015,Bonati2016,Borsanyi2016,Ballesteros2017,Dine2017}, corresponding to Compton frequencies of approximately 0.24--24~GHz. The predicted couplings to axions are model-dependent, though two benchmark models are often used: Kim-Shifman-Vainshtein-Zakharov (KSVZ)~\cite{Kim1979,SHIFMAN1980} and the more stringent Dine-Fischler-Srednicki-Zhitnitsky (DFSZ)~\cite{Zhitnitsky1980,DINE1981}.

\begin{figure}[h]
    \centering
   \includegraphics[width=1.0\columnwidth]{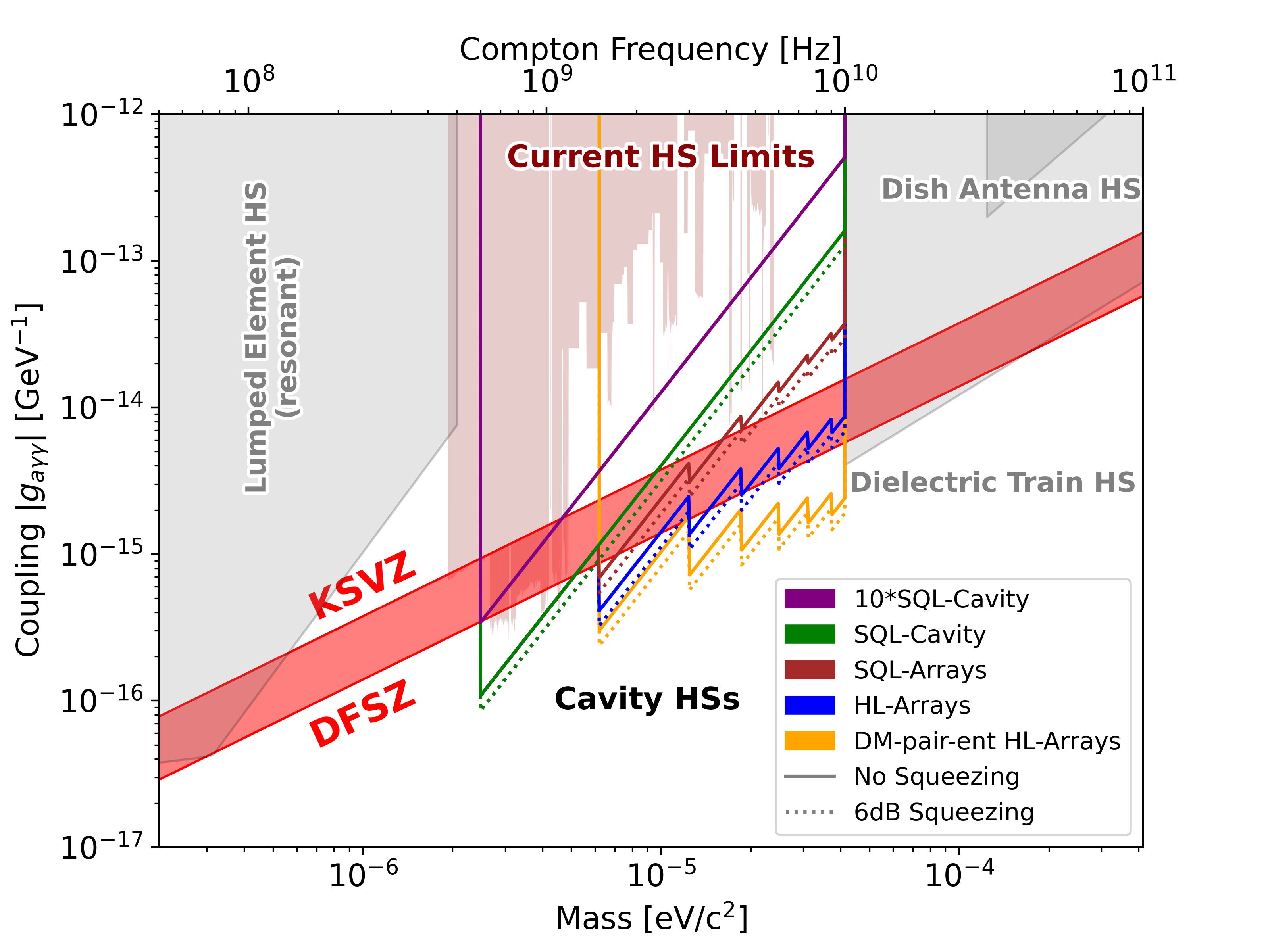}
\caption{Haloscope sensitivity estimates for single cavities (purple and green), coherent cavity arrays (brown), traditional HL cavity arrays (blue), and cavity array with HL readout sensing pair-entangled DM (orange). Dotted lines show enhancement of 6~dB from squeezed readout plus DM squeezing. Existing haloscope limits are in dark red in the background. A cavity has Compton volume $V_{cav} = \lambda_C^3$ and $Q_L = 50,000 (1~\text{GHz}/\nu)^{2/3}$. Teeth of the sawtooth shape represent arrays of size 1, 8, 27, 64, 125, and 343 cavities, read left to right, at increments so that array fits in the original bore size. Note the array volume at 10~GHz is less than 8\% the cavity volume at 600~MHz. Projections for different haloscope techniques in adjacent bands are displayed in gray. Lumped Element HS (DMRadio) is given $Q=100,000$ and volume 500~liters below 80~MHz and the Compton volume $V=\lambda_C^3$ above~\cite{Brouwer_2022}. Dish Antenna HS (BREAD) has 10~m$^2$ collecting area~\cite{BREAD2024}. Dielectric Train HS (MADMAX) has 80 disks of 1~m$^2$ area and boost factor taken from \cite{MADMAX2020Status}. All projections assume $B=10$~Tesla, SQL readout noise scaling, and 10 years total observation time unless otherwise stated. The benchmark KSVZ and DFSZ models coupling strengths are given by the so-labeled red band.
}
\label{fig:haloscope_enhancement}
\end{figure}

\begin{figure}[h]
    \centering
    \includegraphics[width=1.0\columnwidth]{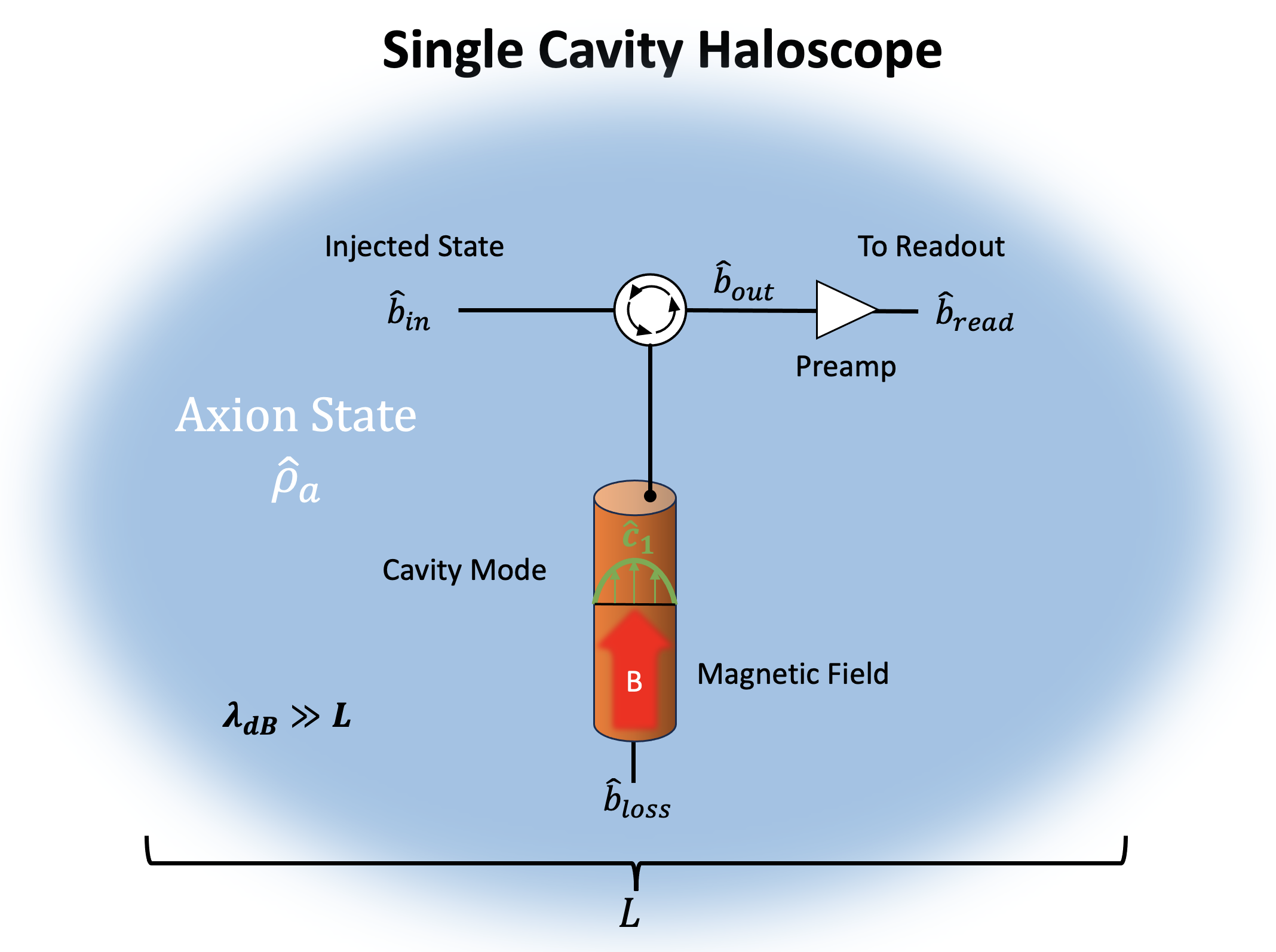}
    \caption{Illustration of a single cavity axion haloscope. }
    \label{fig:cavityqed}
\end{figure}

The search for axion DM in this mass range has been led by cavity haloscopes, a concept originally proposed by Pierre Sikive~\cite{Sikivie1983} that uses a ubiquitous effective coupling between axions and electromagnetism, represented classically by the interaction Lagrange density $\mathscr{L}_{a \gamma \gamma} = \frac{1}{4} g_{a \gamma \gamma} \varphi F_{\alpha \beta} \tilde{F}^{\alpha \beta}  = g_{a \gamma \gamma} \varphi \vec{E} \cdot \vec{B}$ where $\varphi$ is the pseudo-scalar axion field. The axion-photon coupling $g_{a \gamma \gamma}$ is considered to be incredibly feeble, see Fig.~\ref{fig:haloscope_enhancement} for the benchmark models projected into axion mass and axion-photon coupling parameter space over this preferred mass range. The cavity haloscope technique uses this feeble interaction to detect axion DM in a cold microwave cavity threaded by a static magnetic field, with the axion field driving a coupled cavity mode tuned to its frequencies, which is then transmitted by an antenna to the readout circuit. Illustration of a cavity haloscope can be found in Fig.~\ref{fig:cavityqed}. The expected power deposited into a single-cavity haloscope from a monotone classical axion source and averaged over time periods much longer than the coherence time of the cavity mode is 
\begin{equation*}
\begin{aligned}
    &\left< P_{\text{cav}} \right>(\nu) = \frac{\epsilon_0 \alpha^2 c^2}{\pi^2 f_a^2} g_{\gamma}^2 \nu \left< |\varphi|^2 \right>(\nu)  V |B|_{\text{max}}^2 C_{\text{mode}} Q_L T_{\nu_0} (\nu) \\
    &\approx 9.7 \times 10^{-23} W \left(\frac{g_{\gamma}}{0.36}\right)^2 \left(\frac{\rho_a}{0.45~\text{GeV/cm}^{3}}\right) \left(\frac{V}{100~\text{L}}\right) \nonumber \\
    &\left(\frac{B_{\text{max}}}{10~\text{T}}\right)^2 \left(\frac{C_{\text{mode}}}{0.5}\right) \left(\frac{\nu}{1~\text{GHz}}\right) \left(\frac{Q_L}{ 10^5}\right) T_{\nu_0} (\nu) \label{eqn:axionpwr}. 
\end{aligned}
\end{equation*}
Here $g_{\gamma}$ is the unitless coupling strength, $f_a$ is the Peccei-Quinn symmetry breaking energy scale, $\rho_a$ is the local axion DM density, $\epsilon_0$ is the permittivity of free space, $c$ is the speed of light in vacuum, $\alpha$ is the fine structure constant, $V$ is the cavity volume, $B_{\text{max}}$ is the maximum value of the applied magnetic field $Q_L$ is the loaded quality factor of the target cavity mode that has form factor $C_{\text{mode}}$ quantifying the electric and magnetic fields' alignment $C_{\text{mode}} = ( \int_V d^3x \vec{E}_{\text{mode}} \cdot \vec{B} )^2 / (\int_V d^3x |\vec{E}_{\text{mode}}|^2 |B_{\text{max}}|^2 V)$, and $T_{\nu_0} (\nu)$ is the mode envelope shape, expected to be of Lorentzian form. 

It has been the overwhelming prescription in the direct and indirect detection community to describe axion DM as a classical field, or in quantum language as a coherent state\cite{Turner1990,Brubaker2017,Du:2018uak,Braine2020,Bartram2021,ahn2024extensivesearchaxiondark,Goodman2025,Marsh:2015xka,Lentz2019,Lentz2020,Niemeyer_2020,Ferreira_2021,Kopp2022,Eberhardt2022,jaeckel2022reporttopicalgroupwave}. As a coherent state, measurements of the field state are un-impacted by other nearby detectors, with any observed correlation being due to extensive properties of the field's occupied states, such as wavelength. Further, the occupation of these states expected in our neighborhood of the Milky Way (MW), on the order of the virial de Broglie volume $n_{dB} = \rho_a \lambda_{dB}^3/m_a \sim 10^{28} (\uev/m_{DM})^4(300~km \cdot s^{-1}/v_{rms})^3$, are used to motivate simplification of the DM description to the mean field only. This picture has played a significant role in shaping axion searches.

As a result, measuring quantum features in axion DM has been far less considered by the direct detection community. This article seeks to show non-classical axion searches are ripe with opportunity to greatly enhance search sensitivities and uncover novel structures potentially present within the axion DM state. This complements an expanding list of plausible quantum behaviors of the axion state entering the literature~\cite{Duffy2008,Erken2012,Banik2013,Banik2015b,Banik2016,Lentz2019,Lentz2020,Lentz2020b,Lentz2020c,Kopp2022,Eberhardt2022}, including local features such as modified and narrower energy spectra, squeezing, and highly entangled many-body states. Further, recent observations of early cosmological structure formation, cosmic web, galaxies, and galactic substructure have indicated that DM may be collapsing to form halos and other structures faster, denser, and with more diversity than is expected by the Concordance model~\cite{Oman2015,Meneghetti2020,Minor2021,Gilman2023,Xiao2024,roberts2024,Gilman_2025,Zhang_2025,Fei_2025}, many of which have also been hinted at by the above quantum axions studies. It behooves the axion DM community to more closely study the possibility of quantum features in the DM and zero in on the axion state. The goal of this paper is to act on the direct detection facet of this task, demonstrating the potential for signatures of non-classical qualities in the axion DM state to enhance example cavity haloscope searches.

The remainder of this paper is structured as follows. 
Section~\ref{sec:modeling} constructs a dynamical cavity QED (cQED) model of cavity haloscopes under typical operating conditions. 
Section~\ref{sec:examples} applies the cQED model to two classes of axion DM models holding quantum features, revealing candidate non-classical signatures of axion DM in haloscopes. 
Section~\ref{sec:discussion} discusses the impacts of the example signature readout on fiducial cavity haloscope sensitivity, and proposed next steps to realize such non-classical signature searches.

\section{The Quantum Axion Cavity Haloscope} %Fully
\label{sec:modeling}

This section presents a model of quantum axion cavity haloscopes. The model is constructed by first considering an isolated system subsuming the haloscope and DM, then deriving its master equation of motion. The isolated system is then decomposed into the relevant haloscope and axion subsystems, with their equations to follow.

Isolated systems are those comprehensive enough to exhibit symmetry, integrity of the system under transformation of its components~\cite{Courant1953}. In the theory of isolated quantum systems, symmetry is taken to mean unitary transformation of the system's Hilbert space representation $\ket{S} \to \ket{S_U} = \hat{U} \ket{S}$, leaving overall norm of the state (probability amplitude under the Born rule) unchanged $\braket{S_U | S_U} = \bra{S} \hat{U}^{\dagger} \hat{U} \ket{S} = \braket{S | S} = 1$. Time translation symmetry and its generation, the hallmark of isolated physical systems, is the core of this section. The unitarity of this transformation simplifies the derivation of so-called equations of motion for both state vectors and operators within the isolated system according to the linear map $\hat{O} \to \hat{U}^{\dagger} \hat{O} \hat{U}$. The time translation operator may be put into an exponential form, generated by the system's Hamiltonian
\begin{equation}
    \hat{U}(t, t_0) = \exp \left(-i\int_{t_0}^t dt' \hat{H}_{\text{sys}}(t') / \hbar \right). \label{eqn:evolop}
\end{equation}
The translation operator may be decomposed to first order in time to good approximation for small translations ($\delta t \ll \hbar/\hat{H}_{\text{sys}}(t_0)$)
\begin{equation}
    \hat{U}(t_0 + \delta t, t_0) = 1 -i \hat{H}_{\text{sys}}(t_0) \delta t/ \hbar + O(\delta t^2).
\end{equation}
In the calculation of observables, motion may be captured by the operator (Heisenberg picture), the state vector (Schr\"odinger picture), or both (e.g. interaction or Dirac picture). This work will proceed with the Heisenberg picture. The differential equation of motion for an operator in the isolated system is then found to be
\begin{equation}
    \dot{\hat{O}} = \frac{1}{i \hbar} \left[\hat{H}_{\text{sys}}(t), \hat{O}\right]. \label{eqn:opeom}
\end{equation}
When considering the density matrix operator of the system $\hat{\rho}_{\text{sys}}$, the equation of motion for that region comes into view as the von Neumann master equation~\cite{Neumann1927}
\begin{equation}
\dot{\hat{\rho}}_{\text{sys}} = \frac{1}{i \hbar} \left[\hat{H}_{\text{sys}}(t), \hat{\rho}_{\text{sys}}\right]. \label{eqn:vonNeumann}
\end{equation}
The density matrix is defined as $\hat{\rho}_{\text{sys}} = \sum_{\Gamma \in A }p_{\Gamma} \ket{\Gamma} \bra{\Gamma}$ where $p_{\Gamma}$ gives the probability of occupying the pure state $\ket{\Gamma}$ and $A$ is the basis set of the isolated system's Hilbert space.

The collective system considered here includes many constituents from multiple sectors of physics (Standard Model, gravity, the DM sector) organized into many components such as DM halos, cavities, magnets, etc. The components of most interest to this study are the axion DM, the cavity cell modes, and the cavities' receivers. These components will be tracked at the quantum system level. Other coupled components, such as the cavities' walls and magnets, will be treated with effective bulk models. Components deemed inconsequential to the haloscope will be largely omitted.

A frame of reference is not globally well defined in a non-flat space--time in general~\citep{MTW,Wald1984}, and in the presence of gravity some care must be taken in describing a reference frame from which to parameterize the system's time. The extent of a cavity haloscope observation is fortunately small enough, typically in the range of milli-seconds to minutes in time, that a frame centered around the haloscope and evaluated in the Newtonian gauge~\citep{MTW} will suffice. 
The time-like direction that sets the one-parameter family of time evolution is then given by the one-form $\mathbf{n}^* = \sqrt{1 + 2 \Phi} dt$ where $\Phi$ is the weak Newtonian gravitational potential.

The classical Hamiltonian of the system can therefore be found from the system stress-energy $H_{\text{sys}} = T_{\text{sys}}^{\mu\nu} n_{\mu} n_{\nu}$ which itself may be computed from the system effective action functional. The system Hamiltonian for the haloscope is considered to be decomposed into dominant subsystem and subdominant interaction terms
\begin{equation}
    \hat{H}_{\text{sys}} = \sum_i \hat{H}_{i}^0 + \sum_{i, j} \hat{H}_{ij}^{int},
\end{equation}
with associated Hilbert space initially decomposed into component subspaces
\begin{equation}
    \Theta_{\text{sys}} = \Theta_{\text{axion}} \otimes \Theta_{\text{modes}} \Theta_{\text{rec}} \otimes \Theta_{\text{env}},
\end{equation}
where the label ``env'' is used to capture the haloscope environment within the isolated system. Reduced density operators are found by tracing over the complementary Hilbert space $\hat{\rho}_{\text{sub}} = Tr_{\text{sub}^\text{C}} \hat{\rho}_{\text{sys}}$ and are density operators themselves. The subscript $\text{sub}^\text{C}$ indicates the subsystem's complement.  
To derive the form of the equation of motion of a subsystem, trace over the complimentary degrees in the full von Neumann equation
\begin{equation}
    \dot{\hat{\rho}}_{\text{sub}} = 
    \frac{1}{i \hbar } \left[\hat{H}^0_{\text{sub}},\hat{\rho}_{\text{sub}} \right]  
    + \frac{1}{i \hbar } \sum_{j \ne \text{sub}}  Tr_{sub^C} \left[ \hat{H}^{int}_{j,\text{sub}}, \hat{\rho}_{\text{sys}} \right].
\end{equation}

Contributions to the system Hamiltonian for the cavity haloscope will be decomposed effectively as follows
\begin{equation}
    \hat{H}_{\text{sys}} = \hat{H}^0_{\text{axion}} + \hat{H}^0_{\text{mode}} + \hat{H}^0_{\text{rec}} + \hat{H}^{int}_{a \gamma \gamma} + \hat{H}^{int}_{\text{mode}} + \hat{H}^{int}_{\text{env}}.
\end{equation}
The bare axion DM Hamiltonian $\hat{H}^0_{axion}$ is derived from the effective bare classical axion Lagrange density evaluated well below the Peccei-Quinn symmetry breaking and QCD-condensation scales $\mathscr{L}_{\text{axion}} = \frac{1}{2} \nabla_{\mu} \varphi \nabla_{\nu} \varphi g^{\mu \nu} - \frac{1}{2} m^2 \varphi^2$ where the two-tensor $\mathbf{g}$ is the space--time metric. The derived quantum operator form of this Hamiltonian under the assumption of weak perturbing gravity and the Newtonian gauge choice is given by
\begin{align}
    &\hat{H}^0_{\text{axion}} = \\ \nonumber 
    &\frac{1}{2} \int d^3x  \left( m_a^2 \hat{\varphi}^2 - \left(1 + 2 \hat{\Phi} \right) |\vec{\nabla} \hat{\varphi} |^2 + \left(1 - 2 \hat{\Phi} \right) \left(\partial_{0} \hat{\varphi} \right)^2 \right),
\end{align}
where $\hat{\Phi}$ is the weak Newtonian gravitational potential operator. The naive timescale of gravitational infall towards a mass over-density $\delta \rho$ being the turn-around $\tau_g \sim (4 \pi G_N \delta \rho/3)^{-1/2}$ is considered to be too slow compared to the timescales of typical haloscope observations and will be ignored, resulting in a bare free field. The axion field operator $\hat{\varphi}$ is left unexpanded until Section~\ref{sec:examples} when the details of the axion DM models are specified.

The magnetic field induced inside the cavity cells is dominated by sources external to the cells, typically dominated by a superconducting current source. The nature of the source current and induced magnetic field will be taken as coherent states and described using classical fields. The magnetic field from these external sources $\mathbf{B}_0(\mathbf{x})$ will also be considered as time independent in the reference frame of the haloscope, and may be computed from the classical currents using the Biot-Savart law.

The relevant stress energy in a cavity cell sans axion is composed of EM fields within the cavity volume and in the boundary layer, plus the energy contribution of the media in the boundary, where the normal standing modes dominated by EM energy of the cavity cell are of chief interest. The boundary of a cavity cell is considered to be opaque enough to effectively isolate the mode fields between the interior and exterior of a cavity cell, or between cells if multiple cavities exist. The extent of the mode into the boundary material is non-trivial but considered shallow compared to the boundary thickness. The boundary of a cell will be described by multiple channels, including receiver, weak inter-cavity couplings, and loss. The character of the mode is in turn impacted by these boundary interactions. 

The sequence of bare mode wavefunctions can be found by solving the media-Maxwell equations within each cell, forming a Hilbert space for those cell-modes. While each cell contains many mode waveforms, this study is interested in only a single mode per cell that produces a near optimal interaction with the ambient axion field $\varphi$ and static magnetic field $\mathbf{B}_0$, much like the TM010-like modes used in many contemporary cavity haloscopes~\cite{Du:2018uak, Boutan:2017oxg,Brubaker2017,Zhong2018,Braine2020,Backes_2021,Jewell2023,Bai2024,ahn2024extensivesearchaxiondark,Goodman2025}. Mode interference (i.e. mode crossing or hybridization) due to the geometry of the cell is not considered here. The state of these bare TM010-like modes across the cells form a Hilbert space of the form $\Theta_{\text{mode}} = \Theta_{C_1,m} \otimes \Theta_{C_2,m} ... \otimes \Theta_{C_{N_c},m}$ and have the Hamiltonian 
\begin{align}
    \hat{H}^0_{\text{mode}} = \sum_{i}^{N_c} \hbar \omega_{i,m} \left( \hat{c}_{i,m} \hat{c}^{\dagger}_{i,m}   + \frac{1}{2} \right) ,
\end{align}
where $\omega_{i,m}$ is the mode angular frequency for cell $C_i$, and $\hat{c}_{i,m}$, $\hat{c}^{\dagger}_{i,m} $ are the mode's ladder operators. The bare TM010-like mode ladder operators obey the Bosonic commutation relations $[ \hat{c}_{i}, \hat{c}_{j} ] = [ \hat{c}^{\dagger}_{i}, \hat{c}^{\dagger}_{j} ] = 0 $, $[ \hat{c}_{i}, \hat{c}^{\dagger}_{j} ] = \delta_{i j}$.

The bulk systems of the cavity boundary and receiver devices are too extensive and complex to track fully, so this work takes the assumption that the bulk objects are in local near equilibrium. The tracked dynamics will be of standing wave harmonic modes from collective motion of the bulk selected from the near-continuum~\cite{Vool2017,Flower2022thesis}. No material-caused band gaps are considered, only geometry-based gaps. The Hamiltonian for such modes are of the form of a harmonic oscillator $\hat{H}_{b} = \hbar \omega_b  (\hat{b}^{\dagger} \hat{b} + 1/2)$
where the ladder operators $\hat{b}$, $\hat{b}^{\dagger}$ follow Bose commutation relations. The spatial wavefunction of bulk modes will be presented as needed. The interactions between components will be limited to axion-mode-magnetic-field, mode-bulk, mode-mode, and mode-receiver couplings.

The mode-bulk, mode-mode, and mode-receiver interactions are taken to have the same form, consistent with a decohering interaction~\cite{carmichael2010statistical,carmichael2013statistical}
\begin{equation}
    \hat{H}^{int}_{\text{dcoh},ij} = \lambda_{i j} \left( \hat{b}_{i}(t) + \hat{b}^{\dagger}_{i}(t) \right) \left( \hat{b}_{j}(t) + \hat{b}^{\dagger}_{j}(t) \right),
\end{equation}
and will be referred to categorically as $\hat{H}^{int}_{m-b}$, $\hat{H}^{int}_{m-m}$, and $\hat{H}^{int}_{m-r}$ respectively. The decoherence interaction Hamiltonian is bi-linear in the dynamic terms of the relevant subystems, describing the loss of fidelity and introduction of noise as a rotation between subsystems. 

The axion-photon effective interaction couples the ambient axion and magnetic fields in the locales of the cavity modes, with contributing Hamiltonian operator expanded to leading order in magnetic field as 
\begin{align}
    &\hat{H}^{int}_{a \gamma \gamma} = \nonumber g_{a \gamma \gamma} \sum_{i}^{N_{c}} i \sqrt{\frac{ \hbar \omega_i}{2 \epsilon V_i}} \times\\
    & \int d^3x \hat{\varphi}(\mathbf{x},t) \left( \boldsymbol\psi_i (\mathbf{x},t) \hat{c}_i -  \boldsymbol\psi_i^* (\mathbf{x},t) \hat{c}_i^{\dagger} \right) \cdot \mathbf{B}_0 (\mathbf{x}) \nonumber \\
    &+ O(\mathbf{E} \cdot \delta \mathbf{B}),
\end{align}
where the normalized electric field mode waveform is $\boldsymbol\psi_{i,m} = \mathbf{E}_{i,m}/\sqrt{\int_V d^3x |\mathbf{E}_{i,m}|^2 }$. 
For the purpose of this work the axion-photon interaction term will be truncated at the order of the applied magnetic field $\mathbf{B}_0$. 

The resulting equations of motion for the haloscope operators of interest, the modes reduced density matrix operator $\hat{\rho}_{\text{mode}}$ and the mode ladder operators $\hat{c}_i$, $\hat{c}_i^{\dagger}$, can now be written. The reduced density matrix evolution equation becomes
\begin{align}
    \dot{\hat{\rho}}_{\text{mode}} &= 
    \frac{1}{i \hbar } \left[\hat{H}^0_{\text{mode}} + \hat{H}^{int}_{m-m},\hat{\rho}_{\text{mode}} \right] \nonumber \\
    &+ \frac{1}{i \hbar } Tr_{\text{axion}} \left[ \hat{H}^{int}_{a \gamma \gamma}, \hat{\rho}_{m-a} \right] \nonumber \\
    &+ \frac{1}{i \hbar } Tr_{\text{rec}} \left[ \hat{H}^{int}_{m-r}, \hat{\rho}_{m-r} \right] \nonumber \\
    &+ \frac{1}{i \hbar } Tr_{\text{bulk}} \left[ \hat{H}^{int}_{m-b}, \hat{\rho}_{m-b} \right],
\end{align}
where the second, third , and fourth lines involve tracing over the axion, receiver, and bulk subspaces respectively. The equations of motion for the mode, receiver, and bulk ladder operators can also now be specified
\begin{align}
    \dot{\hat{o}}_i &= \frac{1}{i \hbar } \left[\hat{H}^0_{\text{mode}} + \hat{H}^{int}_{m-m} + \hat{H}^{int}_{m-r} + \hat{H}^{int}_{m-b} + \hat{H}^{int}_{a \gamma \gamma},\hat{o}_{i} \right], \nonumber \\
    \dot{\hat{o}}_i^{\dagger} &= \frac{1}{i \hbar } \left[\hat{H}^0_{\text{mode}} + \hat{H}^{int}_{m-m} + \hat{H}^{int}_{m-r} + \hat{H}^{int}_{m-b} + \hat{H}^{int}_{a \gamma \gamma},\hat{o}_{i}^{\dagger} \right].
\end{align}

The next section will elaborate in further detail with two examples of haloscopes and non-classical axion DM signatures.

\section{Example Non-Classical Signatures from Axion DM}
\label{sec:examples}

This section considers the possible signatures from two example non-classical axion DM states onto cavity haloscopes using the dynamical model established in the previous section. The first axion DM state will be of a squeezed coherent state similar to that discussed in \citep{Eberhardt2022,Kopp2022}, where a single cavity haloscope may observe a likewise squeezed signature. The second axion DM state will be of a pair-wise entangled BEC as discussed in \citep{Lentz2019,Lentz2020,Lentz2020b}, where a matched cavity array haloscope may observe a pair-entangled signature across the array's axion-coupled modes.

\subsection{Squeezed Coherent Axion DM in a Single Cavity Haloscope}
\label{sec:squeezedDM}

The axion DM field motion in the vicinity of a haloscope is typically expected to be dominated by free-streaming components. The axion field operator $\hat{\varphi}(x,t)$ may then be accurately decomposed into local linear momentum states as 
\begin{equation}
    \hat{\varphi}(x,t) = \int \frac{d^3p}{(2 \pi)^3 \sqrt{2 E_p}} \left( \hat{a}_p(t) e^{-i \mathbf{p} \cdot \mathbf{x}/\hbar} + \hat{a}_p^{\dagger}(t) e^{i \mathbf{p} \cdot \mathbf{x}/\hbar} \right),
\end{equation}
where $E_p$ is the mode's motive energy used as a normalization factor, and the ground states for each momenta are defined as $\hat{a}_p \ket{0} = 0$. 
Coherent and squeezed states can be created in the momentum basis, where the displacement operator is found to be
\begin{equation}
    \hat{D}(\varphi(x,t)) = \exp \left[ \int \frac{d^3p}{(2 \pi)^3} \left( \hat{a}_p^{\dagger}(t) \tilde{\varphi}(p)  - \hat{a}_p(t) \tilde{\varphi}(p)^* \right) \right],
\end{equation}
where $\tilde{\varphi}(p)$ is the Fourier transform of the position-wise displacement. The squeezing operator on these elements can be defined in this basis as
\begin{equation}
    \hat{S}(\xi(x,t)) = \exp \left[ \frac{1}{2} \int \frac{d^3p}{(2 \pi)^3} \left( \hat{a}_p(t)^2 \tilde{\xi}(p)^* - \hat{a}_p^{\dagger}(t)^2 \tilde{\xi}(p) \right) \right],
\end{equation}
where the squeezing too may vary with space and time. The axion DM's squeezed coherent state is given by the composition $\ket{\varphi,\xi} = \hat{S}(\xi(x,t)) \hat{D}(\varphi(x,t)) \ket{0}$. 
The halo models for $\varphi$ and $\xi$ will be presented as needed later in this sub-section.

The sample haloscope configuration used here to observe squeezed coherent axions contains a single cavity, as in the structure presented in Fig.~\ref{fig:cavityqed}. Further, this haloscope will adopt a receiver using squeezed-state preparation and measurement readout similar to that adopted by the HAYSTAC collaboration~\cite{Malnou2019,Backes_2021,Jewell2023,Bai2024}. To simplify the description of the haloscope, let the mode-bulk coupling take the form $\lambda_{m-b} = \sqrt{\kappa /2 \pi}$ where $\kappa$ gives the dissipation rate to/from the bare mode. For simplicity, the coupling of the receiver channel to the mode will be given the same strength, making the mode and receiver critically coupled. There is no mode-mode coupling as this example considers a single cell and only an isolated TM010-like mode within the cell. The response of the critically-coupled resonant mode centered at angular frequency $\omega_m$ with expected width given by the total loaded dissipation rate $\kappa_L = \kappa_{m-b} + \kappa_{m-r} = 2 \kappa$, which for a typical operating cavity haloscope is characterized by a loaded Q-width of $Q_L = \omega_m/\kappa_L \sim 10^4-10^5$. The axion response is prescribed by the DM model, but will be set such that the Compton angular frequency $\omega_C = m_a c^2/ \hbar$ and have a narrower bandwidth, with sampled Q-widths of at least that expected from a classical virial halo $Q_a \ge 10^6$.

The mode-axion coupling can also be simplified given: (1) the axion field's spatial fluctuations are much broader than the haloscope, with de Broglie length scale of at least $\lambda_{dB} \gtrsim  1000\,{\rm meters}  \times (\uev/m_{DM}) \times (300~\text{km s}^{-1}/ \sqrt{\Delta v^2_{DM}})^2$ for a virialized DM halo compared to the cavity cell extent typically on the order of the axion field's Compton length scale $\lambda_{C} \sim 1\,{\rm meter}  \times (\uev/m_{DM})$; (2) the squeezed coherent axion state is on the diagonal of the axion field operator. The mode-axion Hamiltonian is therefore well approximated by 
\begin{equation}
    \hat{H}_{a \gamma \gamma} = i g_{a \gamma \gamma} \sqrt{\frac{C_m\hbar \omega}{2 V \epsilon_0}} |\bar{B}| \left( \hat{c}(t) -  \hat{c}^{\dagger}(t) \right) \hat{\varphi}(t) + O( \boldsymbol\nabla \hat{\varphi} \cdot \delta \mathbf{x}),
\end{equation}
where the axion field has been taken as uniform over the mode wavefunction, $C_{\text{m}}$ is the geometric form factor quantifying the integral electric field and magnetic field alignment at the mode
\begin{equation}
    C_{m} = \frac{ \left( \int_V d^3x \boldsymbol\psi_{m} \cdot \mathbf{B}_0 \right)^2 }{ |\bar{B}|^2 V } \label{eqn:formfactor},
\end{equation}
and $|\bar{B}|$ is conventionally set as the maximum magnetic field magnitude within the cavity.

The evolution of subsystem operators and the state of the haloscope are now described as follows. The continuum bulk and receiver ladder operator evolution are each of the form
\begin{align}
    \dot{\hat{b}}_k(\omega) &= -i \omega_k\hat{b}_k(\omega) - i \sqrt{\frac{\kappa}{2 \pi}} \hat{c}_m , %\\
    % \dot{\hat{b}}_i^{\dagger}(\omega) &= i \omega\hat{b}_i^{\dagger}(\omega) + i \sqrt{\frac{\kappa}{2 \pi}} \hat{c}^{\dagger}_m,
\end{align}
which have integral form solutions of
\begin{align}
    \hat{b}_k(\omega)(t,t_0) &= e^{-i \omega_k (t-t_0)}\hat{b}_k(\omega)(t_0) \nonumber \\
    &- i \sqrt{\frac{\kappa}{2 \pi}} \int_{t_0}^t dt' e^{-i \omega_k (t-t')} \hat{c}_m(t') , \label{eqn:bulkopevol} %\\
    % \hat{b}_i^{\dagger}(\omega)(t,t_0) &= e^{i \omega (t-t_0)}\hat{b}_i^{\dagger}(\omega)(t_0) \nonumber \\
    % &+ i \sqrt{\frac{\kappa}{2 \pi}} \int_{t_0}^t dt' e^{i \omega (t-t')} \hat{c}^{\dagger}_m(t').
\end{align}
where the form of the $\hat{b}^{\dagger}$ operators are straightforward to derive. 
Further, the sum total response of each of these subsystems' operators onto the mode $\int d \omega \hat{b}_k(\omega)$ may be broken down into input $\hat{b}_{k,in}$ and dynamical components as the respective homogeneous and inhomogeneous parts of Eqn.~\ref{eqn:bulkopevol}. This implies the output from the mode into that subsystem may be considered as its complement given the mode energy loss to that subsystem $\hat{b}_{k,out} = \hat{b}_{k,in} - i \sqrt{\kappa} \hat{c}_m$. 

These forms may be substituted into the mode ladder operators' and density matrix's evolution equations, resulting in the following 
\begin{align}
    &\dot{\hat{\rho}}_m = (-i \omega_m -\kappa)\hat{\rho}_m - i\sqrt{\frac{2 \kappa}{\pi}} \left[ \hat{b}^{\dagger}_{in} \hat{c}_m + \hat{b}_{in} \hat{c}^{\dagger}_m, \hat{\rho}_m \right] \nonumber \\
    &+ g_{a \gamma \gamma} \sqrt{\frac{C_m \omega}{2 V \hbar \epsilon_0}} |\bar{B}| \left( [\hat{c}(t),\hat{\rho}_m] - [\hat{c}^{\dagger}(t),\hat{\rho}_m] \right) \bra{\varphi} \hat{\varphi}(t) \ket{\varphi},
\end{align}
where the evolution of the mode ladder operators are also simplified
\begin{align}
    \dot{\hat{c}}_m &= (-i \omega_m -\kappa)\hat{c}_m + i\sqrt{\frac{2 \kappa}{\pi}}  \hat{b}_{in} \nonumber \\
    &+ g_{a \gamma \gamma} \sqrt{\frac{C_m \omega}{2 V \hbar \epsilon_0}} |\bar{B}| \bra{\varphi} \hat{\varphi}(t) \ket{\varphi},
    % \dot{\hat{c}}_m^{\dagger} &= (i \omega_m -\kappa)\hat{c}_m^{\dagger} - g_{a \gamma \gamma} \sqrt{\frac{C_m \omega}{2 V \hbar \epsilon_0}} |\bar{B}| e^{i \omega_m t} \bra{\varphi} \hat{\varphi}(t) \ket{\varphi},
\end{align}
and again the equation of motion for $\hat{c}_m^{\dagger}$ is straightforward to derive. The state preparation and readout conditions are treated as an initial state and an endpoint observation respectively, similar to the approach in~\citep{Brady2022}. As the cavity mode and receiver channel are critically coupled, the state propagating through the line and into the measurement pre-amplifier will be of the same form and magnitude as the mode occupation in the cavity, dictated by the energy loss boundary condition above. The measurement is treated as an ideal quantum-limited linear amplifier such as a parametric amplifier, meaning that it transforms the mode output according to a two-mode squeezing operation 
\begin{align}
    \hat{b}_{read} &= \hat{S}_2^{\dagger} (s_{\text{amp}}) \hat{b}_{out} \hat{S}_2 (s_{\text{amp}}) \nonumber \\
    &= \cosh (|s_{\text{amp}}|) \hat{b}_{out} - e^{i \theta_{\text{amp}}} \sinh(|s_{\text{amp}}|) \hat{d}^{\dagger}
\end{align}
where $s_{\text{amp}}$ is the complex-valued squeezing parameter, $\theta_{\text{amp}}$ is its angular argument, $\hat{d}$, $\hat{d}^{\dagger}$ are the ladder operators for the amplifier ancillary mode, and $\hat{L}^{\dagger} = - e^{i \theta_{\text{amp}}} \sinh(|s_{\text{amp}}|) \hat{d}^{\dagger}$ is categorized here as the added noise operator~\cite{Caves2012}. Ideally, the ancillary mode will occupy the ground state, giving $\hat{L}$ an expectation of zero and uncertainty $\braket{|\Delta \hat{L}|^2} = 1/2 |\braket{ [ \hat{L}^{\dagger}, \hat{L} ]}| = 1/2\sinh(|s_{\text{amp}}|)^2$, adding noise in quadrature with the amplified input $\cosh (|s_{\text{amp}}|) \hat{b}_{out}$.

\begin{figure}[h!]
    \centering
    \includegraphics[width=0.8\columnwidth]{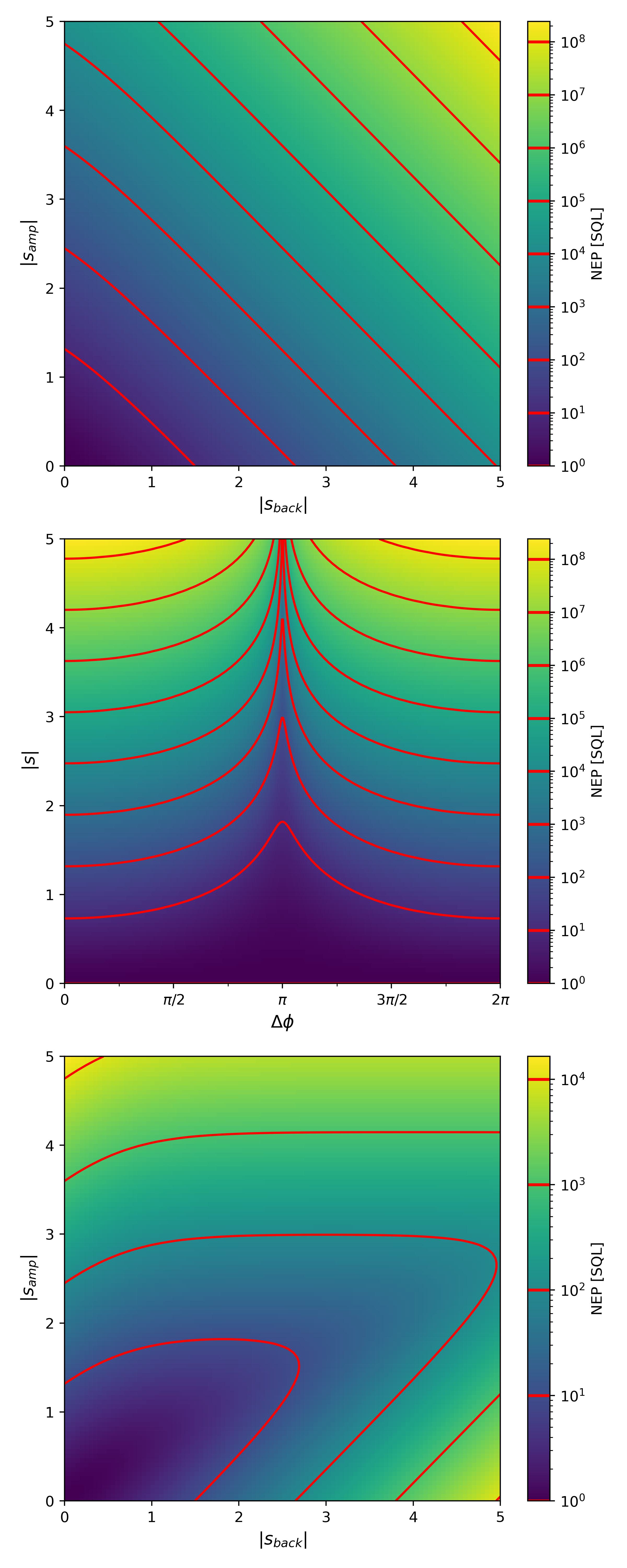}
    \caption{Noise measurements for squeezed state preparation and measurement steps when (upper) the phases are matched $\Delta \phi = 0$ and the preparation and measurement magnitudes are varied, (middle) the squeezing magnitudes are matched and the relative phase is varied, and (lower) the phases are opposite $\Delta \phi = \pi$ and the magnitudes are matched. All noise measurements are made relative to the SQL.}
    \label{fig:sqzNEP}
\end{figure}

The signal-to-noise ratio (SNR) is measured for a chosen observable operator $\hat{O}$ via $SNR = \braket{ \hat{O} }{}^2/\braket{(\Delta \hat{O})^2}$ with the numerator indicating the observable's persistent signal strength and the denominator providing the observable's uncertainty, or noise. The observable operators for this single cavity haloscope will be the quadrature operators of the receiver's measurement port $\hat{Q}_m = 1/\sqrt{2} (\hat{b}_{read}^{\dagger} + \hat{b}_{read})$, $\hat{P}_m = i/\sqrt{2} (\hat{b}_{read}^{\dagger} - \hat{b}_{read})$. For simplicity, this work will consider the SNR based on the quadrature operators' distance from the origin
\begin{equation}
    SNR_m = \frac{\braket{\hat{Q}_m}^2 + \braket{\hat{P}_m}^2 }{\braket{(\Delta \hat{Q}_m)^2 + (\Delta \hat{P}_m)^2}},
\end{equation}
and will be operated as a homodyne detector, considering the integrated observations relative to the mode center frequency. The measurement noise on the single cavity haloscope over different state and measurement squeezing magnitudes and angles show the precision of tuning needed to introduce minimal additional noise to a measurement, Fig.~\ref{fig:sqzNEP}, with the optimal relative angle between preparation and measurement being  $\Delta \phi = \pi$.

Consider the sensitivity of this haloscope to three different axion DM scenarios: a traditional virialized classical field model, a highly coherent classical field, and the squeezed coherent field. Each of these scenarios are pursued with a receiver maintained with the optimal preparation-measurement phase difference $\Delta \phi = \pi$. For these examples, the axion DM field frequency spectrum will be centered on the cavity mode resonance.

\begin{figure}[h]
    \centering
    \includegraphics[width=1.0\columnwidth]{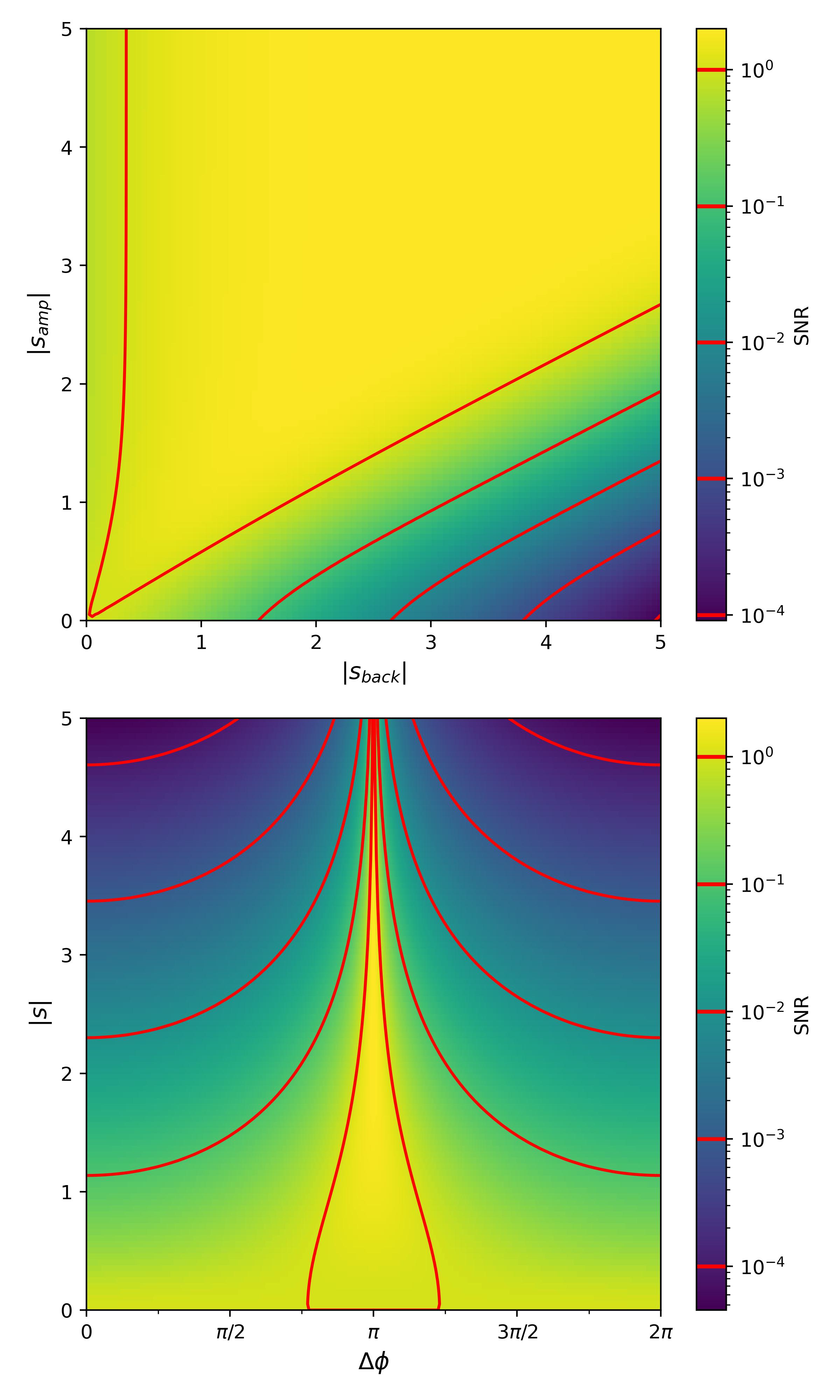}
    \caption{Sections of SNR parameter space of a single haloscope driven by a classical axion field model averaged over many phase domains. All SNRs are relative to the haloscope operating with an unsqueezed SQL receiver. (Top) Background state squeezing magnitude versus amplitude squeezing magnitude, with relative phase of $\Delta \phi = \pi$. (Bottom) Relative phase versus squeezing amplitude $|s| = |s_{amp}| = |s_{back}|$.}
    \label{fig:tradSNR}
\end{figure}

For the traditional classical field model of axion DM~\cite{Turner1990}, the axion field is treated in quantum language as a coherent state with displacement/mean-field $\varphi(\mathbf{x},t)$, which in the vicinity of the single cavity haloscope undergoes drifts in phase randomly on the de Broglie time scale for the virialized halo $\tau_{dB} \sim 4 \times 10^{-3}\,{\rm sec}\times (\uev/m_{DM})\times (300~\text{km s}^{-1}/ \sqrt{\Delta v^2_{DM}})^2$. A typical haloscope observation is expected to persist over many de Broglie times, resulting in a phase-averaged signal. Normalizing the SNR of such a signal processed through a receiver operating at the SQL ($SNR_{SQL} = 1$), Fig.~\ref{fig:tradSNR} displays the regions of sensitivity advantage and disadvantage over the sections of background state squeezing, amplifier squeezing, and their relative phase parameter space.

\begin{figure}[h]
    \centering
    \includegraphics[width=1.0\columnwidth]{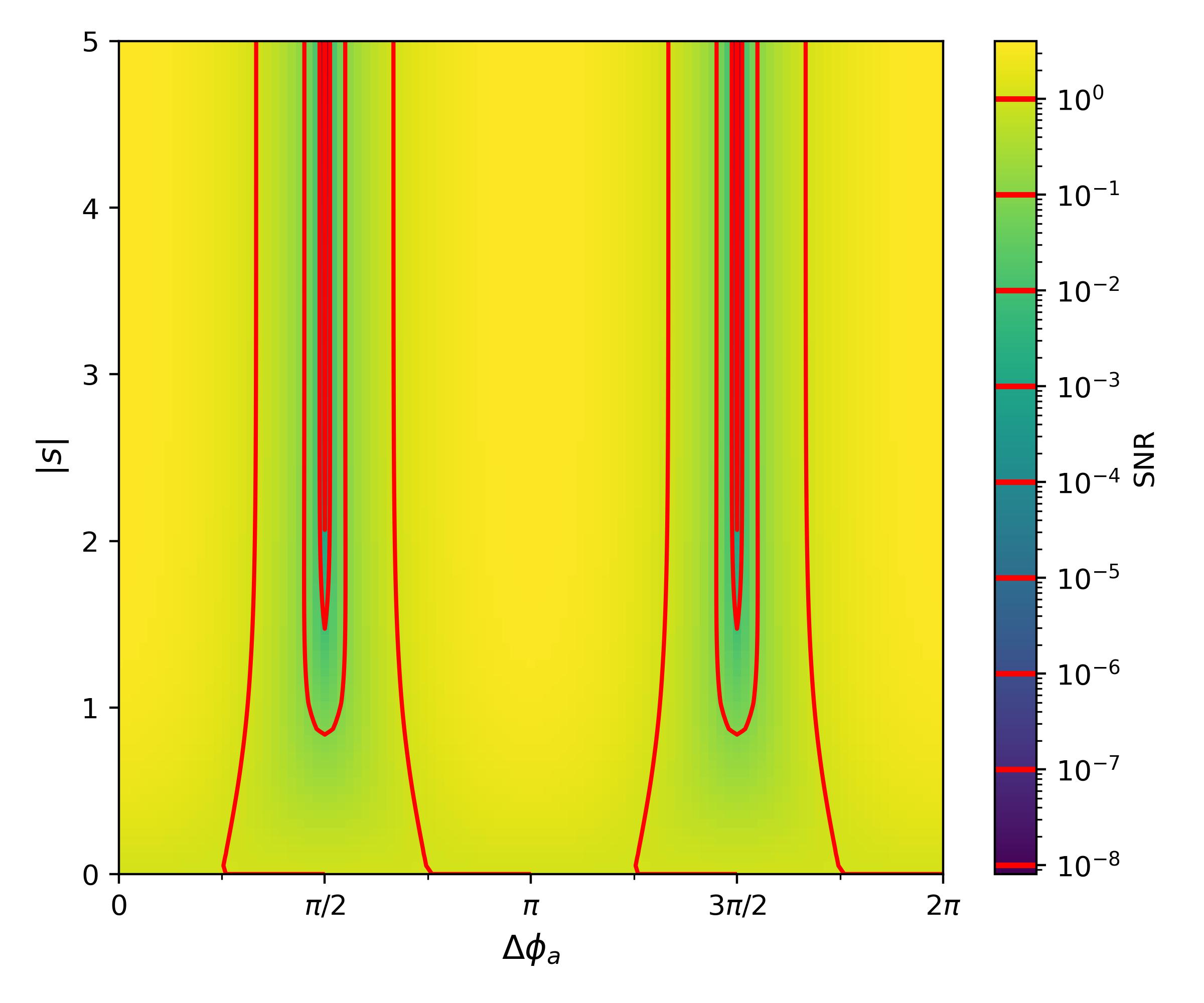}
    \caption{Section of SNR parameter space of an optimally tuned ($\Delta \phi = \pi$) squeezed receiver ($|s| = |s_{amp}| = |s_{back}|$), driven by a coherent signal with its phase set relative to the receiver $\Delta \phi_a$ over the duration of observation.}
    \label{fig:cohSNR}
\end{figure}

A highly coherent axion DM model such as proposed in \citep{Sikivie1999a,Sikivie1999b,Duffy2008,Sikivie2009,Sikivie2011,Erken2012,Banik2013,Banik2015b,Banik2016,Chakrabarty2017} posits the existence of DM cold flows that experience a much smaller degree of velocity dispersion than the traditional virialized halo model $\sqrt{\Delta v^2} \sim 100~\text{km/s} \to \text{cm/s}$, pushing the de Broglie time scale by as many as seven orders of magnitude, possibly beyond the haloscope observation timescale. (Note that there are other structures in the model by~\citep{Sikivie2009}, though this work concentrates only on the coherence of a single local cold flow of DM.) While the local signature is not quantum, the models sourcing such signatures rely on quantum features in the formation of axion halo structure. The relative phase of the receiver to this coherent state now strongly influences the sensitivity of the search, Fig.~\ref{fig:cohSNR}. A further boost in haloscope sensitivity by a factor of two may be experienced if the receiver phase is tuned to match the incoming signal, while a de-tuning may result in a loss of sensitivity in a squeezed receiver. Highly coherent non-virialized axion DM halo models have been the target of haloscope searches~\cite{Duffy2008,Hoskins2016,Bartram2024b,hipp2024searchnonvirializedaxions3342} in the form of ``high-resolution'' heterodyne searches, though not in the context of a squeezed receiver. Such an enhancement must be weighed against the cost of incorporating a sweep over receiver phases into the scanning operations of a haloscope. Considerations of optimal searches for non-classical signatures are beyond the scope of this current work.

\begin{figure}[h]
    \centering
    \includegraphics[width=1.0\columnwidth]{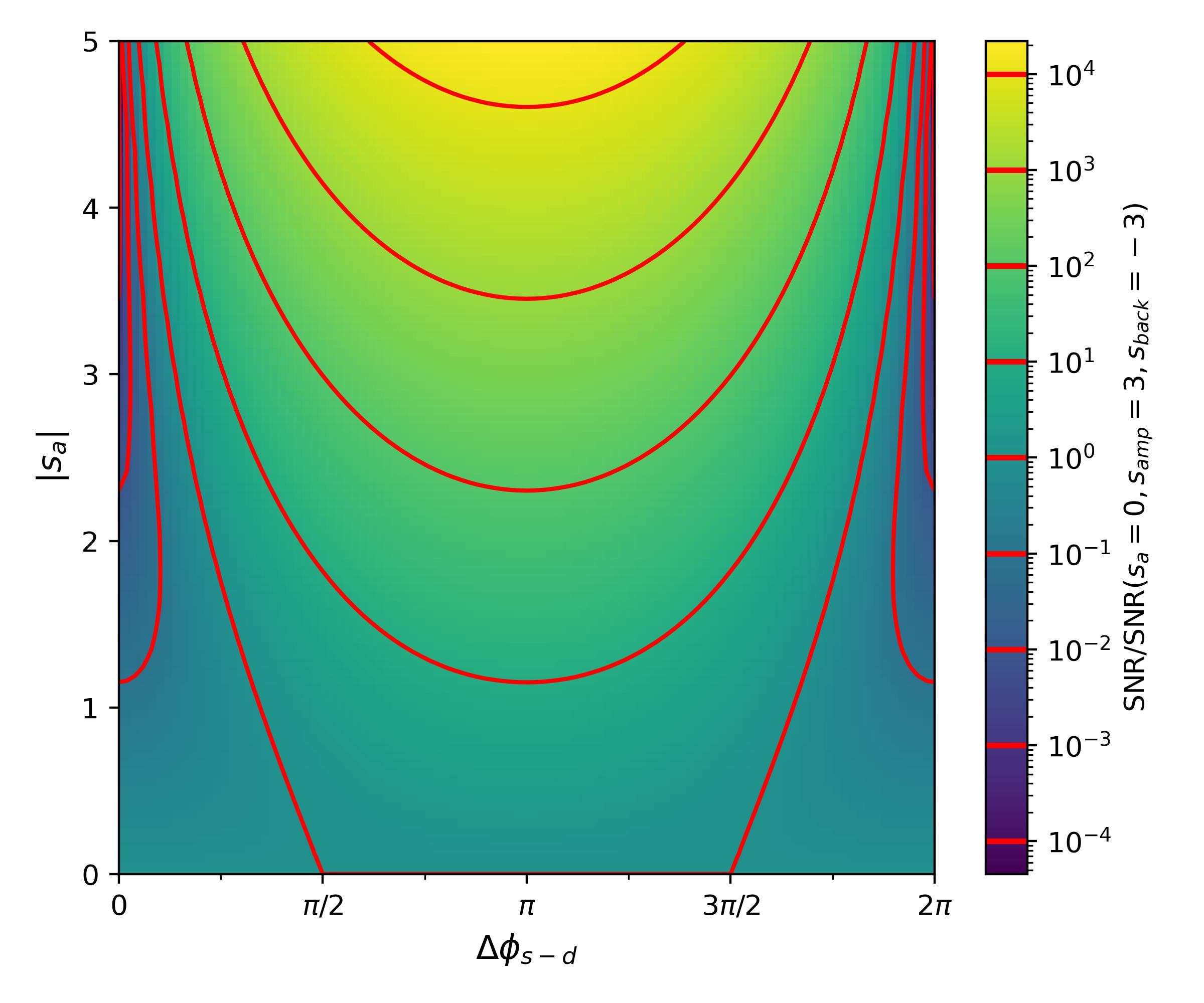}
    \caption{Section of SNR of an optimally tuned ($\Delta \phi = \pi$) receiver with fixed squeezing amplitude ($|s_{amp}| = |s_{back}| = 3$) driven by a squeezed coherent signal. The displacement of the coherent signal is aligned with with receiver ($\Delta \phi_a = 0$) leaving the relative phase of the signal squeezing $\Delta \phi_{s-d}$ free to vary. Phases are held constant over the duration of observation. The magnitude of the signal squeezing factor is allowed to vary $|s_a| \in [0,5]$, though the expected squeezing amplitude is unresolved in the literature~\citep{Eberhardt2022,Kopp2022}.}
    \label{fig:sqzSNR}
\end{figure}

The squeezed model of coherent axion DM introduced at the beginning of this subsection reveals one more opportunity for sensitivity enhancement over the coherent model, related to the DM's squeezing amplitude and alignment. This outcome is fuzzier than the purely coherent model as the additional DM squeezing alters both the signal intensity and the measurement uncertainty, though the prevailing insight of aligning the DM signal's coherence and squeezing phases with the optimal receiver still holds. The haloscope sensitivity over a cross section of the magnitude of the DM squeezing $|s|$ and the relative phase between squeezing and displacement $\Delta \phi_{s-d}$ is provided in Fig.~\ref{fig:sqzSNR}. Scanning techniques for a haloscope seeking a squeezed signal will be unique from that of the only-coherent signal speculated above, most poignantly to marginalize over the unknown relation between the DM's coherence and squeezing.

\subsection{Pair-entangled Axion DM in a Multi-cavity Haloscope}
\label{sec:pairentDM}

The second class of non-classical axion DM models considered here is a BEC of pair-entangled axions of the form
\begin{equation}
    \ket{\Psi} = \ket{\prod_{\alpha \ne \beta} \phi(\mathbf{x}_{\alpha} - \mathbf{x}_{\beta},t)}, \label{eqn:pairentBEC}
\end{equation}
as observed from its center-of-mass frame. This state consists of correlated pairs of spatial operators as presented in \citep{Lentz2019,Lentz2020,Lentz2020b,Lentz2020c}. The length scale of the inter-axion correlation function $\phi(\mathbf{\Delta x},t)$ is expected to be far larger than either the Compton scale or classical virialized halo's de Broglie scale. 

\begin{figure}[h]
    \centering
    \includegraphics[width=1.0\columnwidth]{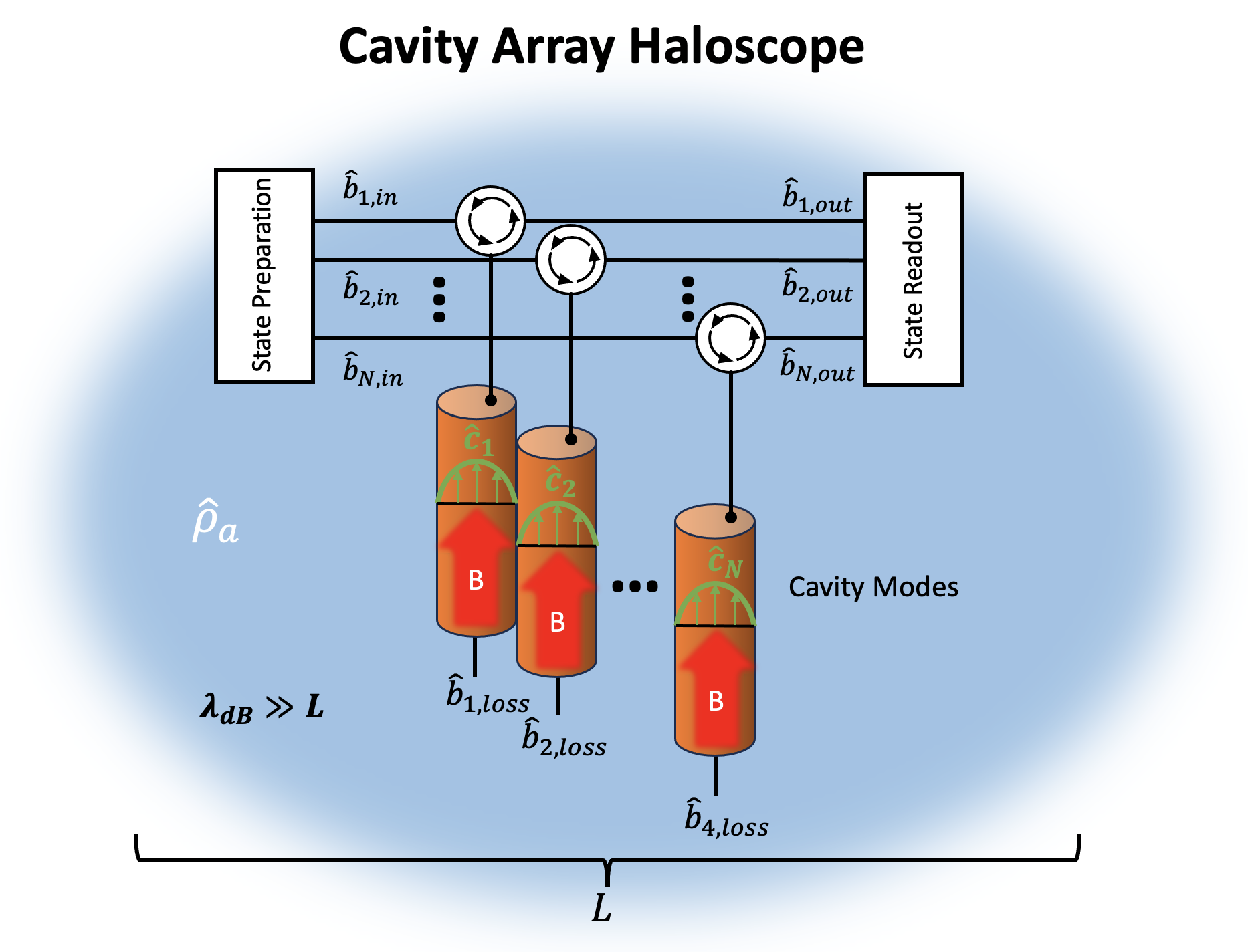}
    \caption{Illustration of a multi-cavity axion haloscope.}
    \label{fig:multicavityarray}
\end{figure}

Such a many-body state may be better observed by a cavity array haloscope instead of a single cavity's mode, which is detailed here. See Fig.~\ref{fig:multicavityarray} for an illustration. Many parameters from the single cavity example will be kept for the arrayed haloscope example
to streamline the presentation, such as matched mode-bulk and mode-receiver couplings, which now apply across all target modes in the array. Each cell in the array will still be considered to have extent on the Compton scale, and that the array will be compact, meaning its total extent will remain much smaller than the virial halo's de Broglie scale $L \ll \lambda_{dB}$.  

The mode-axion Hamiltonian can therefore be well approximated as 
\begin{equation}
    \hat{H}_{a \gamma \gamma} \approx \sum_j^{N_c} i g_{a \gamma \gamma} \sqrt{\frac{C_j \hbar \omega_j}{2 V_j \epsilon_0}} |\bar{B}_j| \left( \hat{c}_j(t) -  \hat{c}_j^{\dagger}(t) \right) \hat{\varphi}_j(t),
\end{equation}
where the axion field operator is now reduced to the domain of each cavity mode 
\begin{equation}
\hat{\varphi}_j = \frac{\int_{V_j} d^3x \hat{\varphi} \boldsymbol\psi_{m} \cdot \mathbf{B}_0}{\int_{V_j} d^3x \boldsymbol\psi_{m} \cdot \mathbf{B}_0}.
\end{equation}
For the remainder of this example the mode angular frequencies, form factors, volumes, and mean magnetic fields will also be considered as matched to a single standard $\omega_j = \omega_m$, $C_j = C_m$, $V_j = V$, $|\bar{B}_j| = |\bar{B}|$.

The evolution equation for the reduced density matrix of the cavity mode is then found to be
\begin{align}
    &\dot{\hat{\rho}}_m = (-i \omega_m -\kappa)\hat{\rho}_m + \sum_{i \ne j}^{N_c} \frac{\lambda_{ij}}{2} \left[\hat{c}_i \hat{c}_j^{\dagger} + \hat{c}_i^{\dagger} \hat{c}_j, \hat{\rho}_m \right]  \nonumber \\
    &- \sum_i^{N_c} i\sqrt{\frac{2 \kappa}{\pi}} \left[ \hat{b}^{\dagger}_{i,in} \hat{c}_i + \hat{b}_{i,in} \hat{c}^{\dagger}_i, \hat{\rho}_m \right] + g_{a \gamma \gamma}  |\bar{B}| \sqrt{\frac{C_m \omega}{2 V \hbar \epsilon_0}} \times \nonumber \\
    &Tr_{\text{axion}} \left(\sum_i^{N_c}  \left( [\hat{c}_i \hat{\varphi}_i,\hat{\rho}_m] - [\hat{c}^{\dagger}_i \hat{\varphi}_i,\hat{\rho}_{m-a}] \right) \right),
\end{align}
where the evolution of the mode ladder operators are of the form
\begin{align}
    \dot{\hat{c}}_i &= (-i \omega_m -\kappa)\hat{c}_i + \sum_{j \ne i}^{N_c} \frac{\lambda_{ij}}{2} \left[\hat{c}_i \hat{c}_j^{\dagger} + \hat{c}_i^{\dagger} \hat{c}_j, \hat{\rho}_m \right] \nonumber \\
    & + i\sqrt{\frac{2 \kappa}{\pi}}  \hat{b}_{i,in} + \sum_j^{N_c}g_{a \gamma \gamma} |\bar{B}| \sqrt{\frac{C_m \omega}{2 V \hbar \epsilon_0}}  [\hat{c}^{\dagger}_j \hat{\varphi}_j,\hat{c}_i],
\end{align}
where again the $\hat{c}_i^{\dagger}$ equations are straightforward to derive. 

The DM model of Eqn.~\ref{eqn:pairentBEC} drives a structurally richer signal into the multi-cavity array. Projected into the individual axion-coupled modes of the cavity array, one finds a signature analogous to the classical field case of Section~\ref{sec:squeezedDM} with amplitude proportional to the local expected axion field $\braket{\hat{Q}_i} \sim \braket{\hat{\varphi}_i}$ and is conducive to detection with modern haloscope receivers. The DM signal also induces inter-mode entanglement via the inter-axion correlations $\phi(\mathbf{\Delta x},t)$. These inter-mode entanglements can be seen in the two-cavity observables, including $\braket{\hat{Q}_i \hat{Q}_j+\hat{P}_i \hat{P}_j}$ where the signature takes the form
\begin{align}
    \braket{\hat{Q}_i \hat{Q}_j+\hat{P}_i \hat{P}_j} &\sim \epsilon g_{a \gamma \gamma}^2 |\bar{B}|^2 V C_m Q_L T_{\nu_0}(\nu) \nu \times \nonumber \\
    &\int d^3x_1 ... d^3x_{N_a} \left( \varphi_i \varphi_j \prod_{a \ne b} \phi^2(\mathbf{x}_a - \mathbf{x}_b) \right) \nonumber \\
    % &= \epsilon g_{a \gamma \gamma}^2 |\bar{B}|^2 V C_m Q_L T_{\nu_0}(\nu) \nu \rho_a \frac{\phi_{ij}^2}{\phi_0^2} \nonumber \\
    &= \braket{P_{\text{cav},i}} \phi_{ij}^2/\phi_{ii}^2, \label{eqn:pairsigstrength}
\end{align}
where
\begin{equation}
    \phi^2_{ij} = \frac{\int d^3x_1 d^3x_2 (\boldsymbol\psi_{i} \cdot \mathbf{B}_0 )(\mathbf{x}_1) \phi^2(\mathbf{x}_1 - \mathbf{x}_2)  (\boldsymbol\psi_{j} \cdot \mathbf{B}_0) (\mathbf{x}_2)}{\int d^3x_1 \boldsymbol\psi_{i}(\mathbf{x}_1) \cdot \mathbf{B}_0(\mathbf{x}_1) \int d^3x_2 \boldsymbol\psi_{j}(\mathbf{x}_2) \cdot \mathbf{B}_0(\mathbf{x}_2)}.
\end{equation}
This two-cavity signature contains both quantum and classical components. Classical correlations between the output of individual cavities in a pair $\braket{\hat{Q}_i} \braket{\hat{Q}_j}$ are produced from the time-wise correlations in the axion mean field between cavity mode sites, $\braket{\hat{\varphi}_i} \braket{\hat{\varphi}_j}$. The novel non-classical correlations between two cavities are given by the inseparable components $\braket{\hat{Q}_i \hat{Q}_j}-\braket{\hat{Q}_i} \braket{\hat{Q}_j}$, generated by the inseparable component of the inter-axion correlation function evaluated at two cavity modes $\braket{\hat{\varphi}_i \hat{\varphi}_j}-\braket{\hat{\varphi}_i} \braket{\hat{\varphi}_j}$. 
Further, it should be noted that isolation of the cavity modes across the array is crucial to the uniqueness of inseparable component of the signature as a coherent signal injected into a cavity array with mode-mode couplings will produce similar inter-mode correlations. The mode-mode couplings are set to zero, $\lambda_{ij} = 0$, for this example. Additional axion signatures do appear in higher-order mode correlators, though this demonstration will be restricted to correlators containing at most two modes, with higher order two-mode correlators considered as reducible to first- and second-order correlators. 

Two different readout types will be compared in this example corresponding to the two signatures highlighted above: a Heisenberg-limited (HL) power combiner intended to observe the single modes' signature~\cite{Brady2022}, and a HL inter-mode correlation combiner intended to observe the inseparable two-mode correlations' signature ~\cite{Verstraete2004,ORUS2014,Cirac2021}. The specific designs of the receivers are left for future work. The sensitivity measure for the combined power readout from the array is
\begin{equation}
    SNR_{\text{pwr}} = \frac{\braket{\hat{Q}_m}^2 + \braket{\hat{P}_m}^2}{\braket{(\Delta \hat{Q}_m)^2 + ( \Delta\hat{P}_m)^2}}, \label{eqn:SNRpwr} 
\end{equation}
where now the measurement operators are a superposition of cavity outputs with ideal weights
($\hat{Q}_m = \sum_i \hat{Q}_i/\sqrt{N_c}$, $\hat{P}_m = \sum_i \hat{P}_i/\sqrt{N_c}$). The sensitivity measure for inter-mode entanglement read out from the same array is taken via projection onto the combined pair-correlator operators and has the form
\begin{equation}
    SNR_{\text{corr}} = \frac{\braket{\widehat{QQ}_{\text{corr}}}^2 + \braket{\widehat{PP}_{\text{corr}}}^2}{\braket{(\Delta\widehat{QQ}_{\text{corr}})^2 + (\Delta\widehat{PP}_{\text{corr}})^2}}, \label{eqn:SNRcorr}
\end{equation}
where the measured correlation operators are an ideal superposition of individual pair correlators $\widehat{QQ}_{\text{corr}} = \binom{N_c}{2}^{-1/2} \sum_{i \ne j} \hat{Q}_i \hat{Q}_j$, $\widehat{PP}_{\text{corr}} = \binom{N_c}{2}^{-1/2} \sum_{i \ne j} \hat{P}_i  \hat{P}_j$. Again, the details and susceptibilities of these receivers are beyond the scope of this paper and deferred to future work in order to concentrate on the properties of the signatures under ideal detection conditions, though the noise of both the power-combining and correlation-combining receivers are assumed to be HL with an adequately clean network and appropriately prepared state~\cite{Verstraete2004,ORUS2014,Cirac2021,Brady2022}. 

\begin{figure}[h]
    \centering
    \includegraphics[width=1.0\columnwidth]{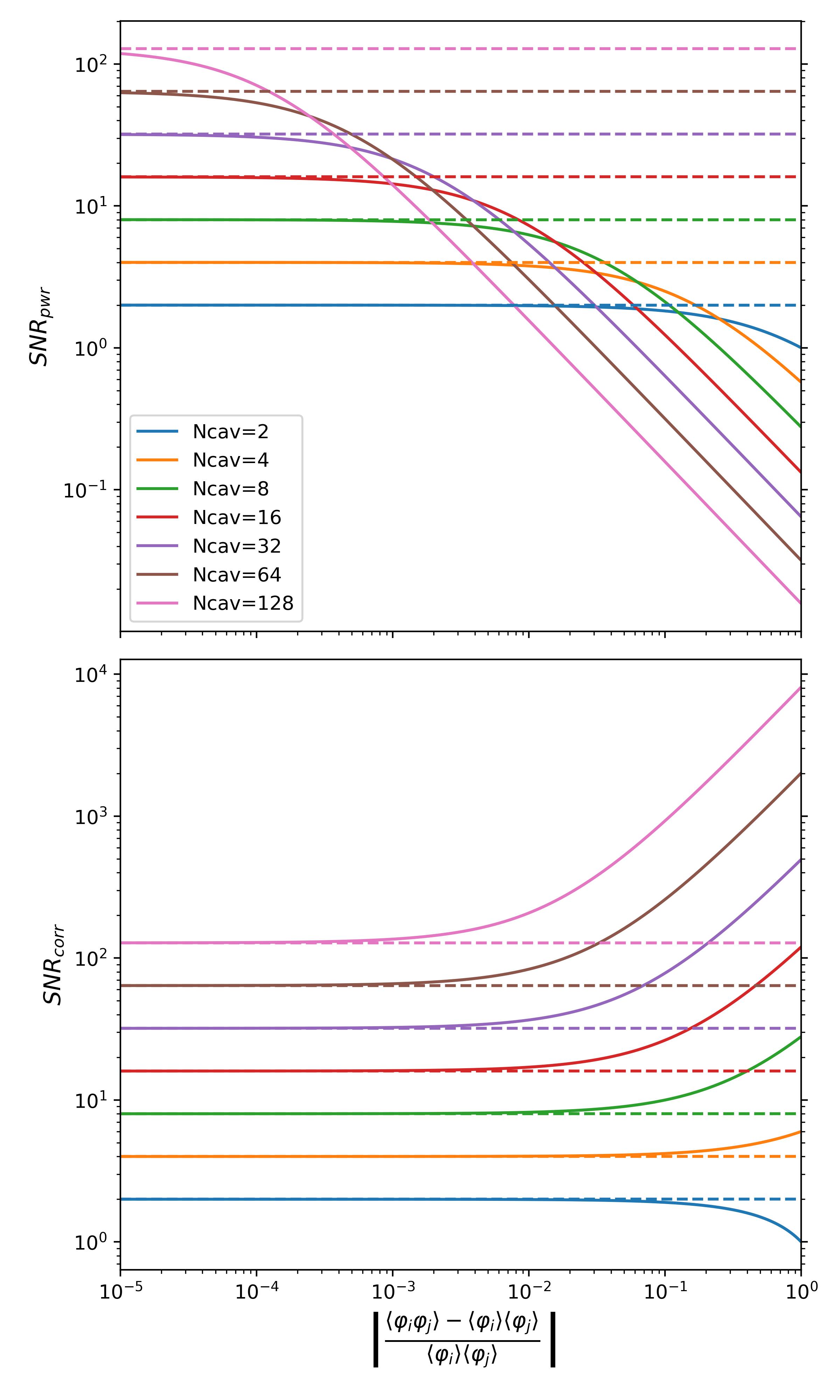}
    \caption{Single observation signal-to-noise scaling estimates over cavity array size and inter-axion correlation strength (solid lines) referenced to a classical coherent field (dashed). (Top) Signal-to-noise scaling of the power combiner receiver. (Bottom) Signal-to-noise scaling of the pair-correlator receiver. Axion signal strength is calibrated to $SNR_{\text{pwr}} = 1$ for a single cavity haloscope.}
    \label{fig:cavarraySNR}
\end{figure}

The strength of the signature from the combined power receiver $SNR_{\text{pwr}}$ nominally scales as the number of cavities in the array $N_c$. Further, the strength of the correlation signature $SNR_{\text{corr}}$ may scale as high as the number of pair correlators being combined, $N_c(N_c-1)/2$, while maintaining measurement uncertainty as low as a single degree of freedom. This provides an opportunity to improve upon the naive HL cavity array for array sizes $N_c > 3$ when the individual pair correlator signatures are of the same strength as the individual cavity output power $\braket{P_{\text{cav}}}$. 
The behavior of both receivers over different array sizes and inter-axion correlation strengths are provided in Fig.~\ref{fig:cavarraySNR}. The power combiner receiver $SNR_{\text{pwr}}$ is seen to depend on the inter-axion correlation strength as the second-order correlations enlarge the measurement variance, having an adverse impact to that haloscope's sensitivity if the feature is not isolated and removed. The receiver attuned to inter-mode entanglement $SNR_{\text{corr}}$ is expectantly highly sensitive to the correlation feature, but degrades when the DM model becomes separable across the cavity array $\braket{\hat{\varphi}_i \hat{\varphi}_j}-\braket{\hat{\varphi}_i} \braket{\hat{\varphi}_j} \to 0$.

\section{Summary}
\label{sec:discussion}

The above examples of non-classical signatures in cavity haloscopes have demonstrated that significant sensitivity enhancement is possible with an appropriately calibrated receiver, or diminished if a poorly calibrated receiver is used. %Two example axion DM models were pursued for their potential non-classical signatures: a squeezed coherent state and a pair-entangled state.
% For well-calibrated receivers, projections of cavity haloscope sensitivity to the example signatures can be made, Fig.~\ref{fig:haloscope_enhancement}. 
In the first of these scenarios, a highly coherent squeezed axion field can notably increase the sensitivity of a cavity haloscope that is itself equipped with a squeezed receiver, so long as that receiver is aligned with the (co-aligned) displacement and squeezing directions of the local DM, thereby extending the potency of a cavity haloscope techniques to reach frequencies well above a GHz while maintaining DFSZ sensitivity. In the second scenario, a pair-entangled Bose condensate state was seen to induce inter-mode entanglement signatures in a cavity array haloscope. If the DM is strongly entangled, the more typical power combing receivers may be adversely impacted by the presence of additional variance unknowingly sourced by the DM itself, though it may be possible to mitigate this impact in the analysis phase.
If using a pair-correlation receiver, this signature has the potential to produce a significant sensitivity scaling enhancement compared a classical DM model. Such a scenario may allow cavity array haloscopes to feasibly maintain DFSZ sensitivity through and beyond the 10~GHz band.

This exciting result is not without its challenges. The next steps to build upon this work are to further develop the haloscope-receiver models and benchmark devices suitable for such haloscope readout systems, growing towards prototype non-classical axion signature searches. This will include the development of modeling frameworks for further detailing the nature and lifetime of signatures when measured through a haloscope's cascade of RF circuit components, subject to measured or projected component transmission properties and sensitivities to environmental conditions (e.g., temperature and magnetic fields), and will be used to to refine the readout design and protocol by tracing the response of the full RF signal chain, including quantum-sensitive elements. Modeling will rotate into physical detector R\&D to fabricate novel component designs and quantum readout systems. Further, the haloscopes outlined in this work are configured for specific non-classical signatures, based on speculative axion DM models. Another fruitful research path that can be pursued in parallel to the haloscope R\&D is the refinement of many-body axion DM modeling such that it can provide realistic models of Milky Way-like halos at galactic and down to haloscope scales.

\section{Acknowledgements}
\label{sec:acknowledgements}

I would like to express my gratitude to Michael Kopp, Benjo Fraser, Andrew Eberhardt, Christian Boutan, and John Orrell for their insightful discussions of this research and manuscript. The research described in this paper was partially supported under the Laboratory Directed Research and Development Program at Pacific Northwest National Laboratory, a multi-program national laboratory operated by Battelle for the U.S. Department of Energy (DOE) under Contract No. DE-AC05-76RL01830. This work was also partially supported by the Advanced Technology Research and Development program within the DOE Office of High Energy Physics as well as a grant from the US Department of Energy Office of High Energy Physics Quantum Information Science Enabled Discovery (QuantISED) program.

\bibliography{Bibliography}

\begin{thebibliography}{10}

\bibitem{P5report2023}
Exploring the quantum universe: Pathways to innovation and discovery in
  particle physics.

\bibitem{ABBOTT1983}
L.F. Abbott and P.~Sikivie.
\newblock A cosmological bound on the invisible axion.
\newblock {\em Physics Letters B}, 120(1):133 -- 136, 1983.

\bibitem{Planck2018}
N.~Aghanim, Y.~Akrami, F.~Arroja, M.~Ashdown, J.~Aumont, C.~Baccigalupi,
  M.~Ballardini, A.~J. Banday, R.~B. Barreiro, N.~Bartolo, S.~Basak, R.~Battye,
  K.~Benabed, J.-P. Bernard, M.~Bersanelli, P.~Bielewicz, J.~J. Bock, J.~R.
  Bond, J.~Borrill, F.~R. Bouchet, F.~Boulanger, M.~Bucher, C.~Burigana, R.~C.
  Butler, E.~Calabrese, J.-F. Cardoso, J.~Carron, B.~Casaponsa, A.~Challinor,
  H.~C. Chiang, L.~P.~L. Colombo, C.~Combet, D.~Contreras, B.~P. Crill,
  F.~Cuttaia, P.~de~Bernardis, G.~de~Zotti, J.~Delabrouille, J.-M. Delouis,
  F.-X. Désert, E.~Di~Valentino, C.~Dickinson, J.~M. Diego, S.~Donzelli,
  O.~Doré, M.~Douspis, A.~Ducout, X.~Dupac, G.~Efstathiou, F.~Elsner, T.~A.
  Enßlin, H.~K. Eriksen, E.~Falgarone, Y.~Fantaye, J.~Fergusson,
  R.~Fernandez-Cobos, F.~Finelli, F.~Forastieri, M.~Frailis, E.~Franceschi,
  A.~Frolov, S.~Galeotta, S.~Galli, K.~Ganga, R.~T. Génova-Santos, M.~Gerbino,
  T.~Ghosh, J.~González-Nuevo, K.~M. Górski, S.~Gratton, A.~Gruppuso, J.~E.
  Gudmundsson, J.~Hamann, W.~Handley, F.~K. Hansen, G.~Helou, D.~Herranz, S.~R.
  Hildebrandt, E.~Hivon, Z.~Huang, A.~H. Jaffe, W.~C. Jones, A.~Karakci,
  E.~Keihänen, R.~Keskitalo, K.~Kiiveri, J.~Kim, T.~S. Kisner, L.~Knox,
  N.~Krachmalnicoff, M.~Kunz, H.~Kurki-Suonio, G.~Lagache, J.-M. Lamarre,
  M.~Langer, A.~Lasenby, M.~Lattanzi, C.~R. Lawrence, M.~Le~Jeune, J.~P. Leahy,
  J.~Lesgourgues, F.~Levrier, A.~Lewis, M.~Liguori, P.~B. Lilje, M.~Lilley,
  V.~Lindholm, M.~López-Caniego, P.~M. Lubin, Y.-Z. Ma, J.~F. Macías-Pérez,
  G.~Maggio, D.~Maino, N.~Mandolesi, A.~Mangilli, A.~Marcos-Caballero,
  M.~Maris, P.~G. Martin, M.~Martinelli, E.~Martínez-González, S.~Matarrese,
  N.~Mauri, J.~D. McEwen, P.~D. Meerburg, P.~R. Meinhold, A.~Melchiorri,
  A.~Mennella, M.~Migliaccio, M.~Millea, S.~Mitra, M.-A. Miville-Deschênes,
  D.~Molinari, A.~Moneti, L.~Montier, G.~Morgante, A.~Moss, S.~Mottet,
  M.~Münchmeyer, P.~Natoli, H.~U. Nørgaard-Nielsen, C.~A. Oxborrow,
  L.~Pagano, D.~Paoletti, B.~Partridge, G.~Patanchon, T.~J. Pearson, M.~Peel,
  H.~V. Peiris, F.~Perrotta, V.~Pettorino, F.~Piacentini, L.~Polastri,
  G.~Polenta, J.-L. Puget, J.~P. Rachen, M.~Reinecke, M.~Remazeilles,
  C.~Renault, A.~Renzi, G.~Rocha, C.~Rosset, G.~Roudier, J.~A. Rubiño-Martín,
  B.~Ruiz-Granados, L.~Salvati, M.~Sandri, M.~Savelainen, D.~Scott, E.~P.~S.
  Shellard, M.~Shiraishi, C.~Sirignano, G.~Sirri, L.~D. Spencer, R.~Sunyaev,
  A.-S. Suur-Uski, J.~A. Tauber, D.~Tavagnacco, M.~Tenti, L.~Terenzi,
  L.~Toffolatti, M.~Tomasi, T.~Trombetti, J.~Valiviita, B.~Van~Tent, L.~Vibert,
  P.~Vielva, F.~Villa, N.~Vittorio, B.~D. Wandelt, I.~K. Wehus, M.~White,
  S.~D.~M. White, A.~Zacchei, and A.~Zonca.
\newblock Planck2018 results: I. overview and the cosmological legacy of
  planck.
\newblock {\em Astronomy \&; Astrophysics}, 641:A1, September 2020.

\bibitem{ahn2024extensivesearchaxiondark}
Saebyeok Ahn, JinMyeong Kim, Boris~I. Ivanov, Ohjoon Kwon, HeeSu Byun, Arjan~F.
  van Loo, SeongTae Park, Junu Jeong, Soohyung Lee, Jinsu Kim, \ifmmode
  \mbox{\c{C}}\else \c{C}\fi{}a\ifmmode \breve{g}\else~\u{g}\fi{}lar Kutlu,
  Andrew~K. Yi, Yasunobu Nakamura, Seonjeong Oh, Danho Ahn, SungJae Bae,
  Hyoungsoon Choi, Jihoon Choi, Yonuk Chong, Woohyun Chung, Violeta Gkika,
  Jihn~E. Kim, Younggeun Kim, Byeong~Rok Ko, Lino Miceli, Doyu Lee, Jiwon Lee,
  Ki~Woong Lee, MyeongJae Lee, Andrei Matlashov, Pallavi Parashar, Taehyeon
  Seong, Yun~Chang Shin, Sergey~V. Uchaikin, SungWoo Youn, and Yannis~K.
  Semertzidis.
\newblock Extensive search for axion dark matter over 1 ghz with capp's main
  axion experiment.
\newblock {\em Phys. Rev. X}, 14:031023, Aug 2024.

\bibitem{Backes_2021}
Kelly~M Backes, Daniel~A Palken, S~Al Kenany, Benjamin~M Brubaker, SB~Cahn,
  A~Droster, Gene~C Hilton, Sumita Ghosh, H~Jackson, Steve~K Lamoreaux, et~al.
\newblock A quantum enhanced search for dark matter axions.
\newblock {\em Nature}, 590(7845):238--242, 2021.

\bibitem{Ballesteros2017}
Guillermo Ballesteros, Javier Redondo, Andreas Ringwald, and Carlos Tamarit.
\newblock Unifying inflation with the axion, dark matter, baryogenesis, and the
  seesaw mechanism.
\newblock {\em Phys. Rev. Lett.}, 118:071802, Feb 2017.

\bibitem{Banik2015b}
N.~{Banik}, A.~J. {Christopherson}, P.~{Sikivie}, and E.~M. {Todarello}.
\newblock {Linear Newtonian perturbation theory from the
  Schr{\"o}dinger-Poisson equations}.
\newblock {\em \prd}, 91(12):123540, June 2015.

\bibitem{Banik2013}
N.~{Banik} and P.~{Sikivie}.
\newblock {Axions and the galactic angular momentum distribution}.
\newblock {\em \prd}, 88(12):123517, December 2013.

\bibitem{Banik2016}
N.~{Banik} and P.~{Sikivie}.
\newblock {Evolution of velocity dispersion along cold collisionless flows}.
\newblock {\em PRD}, 93(10):103509, May 2016.

\bibitem{Bartram2024b}
C.~Bartram, T.~Braine, R.~Cervantes, N.~Crisosto, N.~Du, C.~Goodman,
  M.~Guzzetti, C.~Hanretty, S.~Lee, G.~Leum, L.~J Rosenberg, G.~Rybka,
  J.~Sinnis, D.~Zhang, M.~H. Awida, D.~Bowring, A.~S. Chou, M.~Hollister,
  S.~Knirck, A.~Sonnenschein, W.~Wester, R.~Khatiwada, J.~Brodsky, G.~Carosi,
  L.~D. Duffy, M.~Goryachev, B.~McAllister, A.~Quiskamp, C.~Thomson, M.~E.
  Tobar, C.~Boutan, M.~Jones, B.~H. LaRoque, E.~Lentz, N.~E. Man, N.~S. Oblath,
  M.~S. Taubman, J.~Yang, John Clarke, I.~Siddiqi, A.~Agrawal, A.~V. Dixit,
  J.~R. Gleason, Y.~Han, A.~T. Hipp, S.~Jois, P.~Sikivie, N.~S. Sullivan, D.~B.
  Tanner, E.~J. Daw, M.~G. Perry, J.~H. Buckley, C.~Gaikwad, J.~Hoffman, K.~W.
  Murch, and J.~Russell.
\newblock Nonvirialized axion search sensitive to doppler effects in the milky
  way halo.
\newblock {\em Phys. Rev. D}, 109:083014, Apr 2024.

\bibitem{Bartram2021}
C.~Bartram, T.~Braine, R.~Cervantes, N.~Crisosto, N.~Du, G.~Leum, L.~J.
  Rosenberg, G.~Rybka, J.~Yang, D.~Bowring, A.~S. Chou, R.~Khatiwada,
  A.~Sonnenschein, W.~Wester, G.~Carosi, N.~Woollett, L.~D. Duffy,
  M.~Goryachev, B.~McAllister, M.~E. Tobar, C.~Boutan, M.~Jones, B.~H. LaRoque,
  N.~S. Oblath, M.~S. Taubman, John Clarke, A.~Dove, A.~Eddins, S.~R. O'Kelley,
  S.~Nawaz, I.~Siddiqi, N.~Stevenson, A.~Agrawal, A.~V. Dixit, J.~R. Gleason,
  S.~Jois, P.~Sikivie, J.~A. Solomon, N.~S. Sullivan, D.~B. Tanner, E.~Lentz,
  E.~J. Daw, M.~G. Perry, J.~H. Buckley, P.~M. Harrington, E.~A. Henriksen, and
  K.~W. Murch.
\newblock Axion dark matter experiment: Run 1b analysis details.
\newblock {\em Phys. Rev. D}, 103:032002, Feb 2021.

\bibitem{Berges2015}
J\"urgen Berges and Joerg Jaeckel.
\newblock Far from equilibrium dynamics of bose-einstein condensation for axion
  dark matter.
\newblock {\em Phys. Rev. D}, 91:025020, Jan 2015.

\bibitem{Berkowitz2015}
Evan Berkowitz, Michael~I. Buchoff, and Enrico Rinaldi.
\newblock Lattice qcd input for axion cosmology.
\newblock {\em Phys. Rev. D}, 92:034507, Aug 2015.

\bibitem{MADMAX2020Status}
S.~Beurthey, N.~Böhmer, P.~Brun, A.~Caldwell, L.~Chevalier, C.~Diaconu,
  G.~Dvali, P.~Freire, E.~Garutti, C.~Gooch, A.~Hambarzumjan, S.~Heyminck,
  F.~Hubaut, J.~Jochum, P.~Karst, S.~Khan, D.~Kittlinger, S.~Knirck, M.~Kramer,
  C.~Krieger, T.~Lasserre, C.~Lee, X.~Li, A.~Lindner, B.~Majorovits,
  M.~Matysek, S.~Martens, E.~Öz, P.~Pataguppi, P.~Pralavorio, G.~Raffelt,
  J.~Redondo, O.~Reimann, A.~Ringwald, N.~Roch, K.~Saikawa, J.~Schaffran,
  A.~Schmidt, J.~Schütte-Engel, A.~Sedlak, F.~Steffen, L.~Shtembari,
  C.~Strandhagen, D.~Strom, and G.~Wieching.
\newblock Madmax status report, 2020.

\bibitem{Bonati2016}
Claudio {Bonati}, Massimo {D'Elia}, Marco {Mariti}, Guido {Martinelli}, Michele
  {Mesiti}, Francesco {Negro}, Francesco {Sanfilippo}, and Giovanni
  {Villadoro}.
\newblock {Axion phenomenology and {\ensuremath{\theta}}-dependence from N $_{
  f }$ = 2 + 1 lattice QCD}.
\newblock {\em Journal of High Energy Physics}, 2016:155, March 2016.

\bibitem{Borsanyi2016}
Szabolcs Borsanyi, Zoltán Fodor, J.~Guenther, Karl-Heinz Kampert, S.~Katz,
  T.~Kawanai, Tamás Kovács, S.~Mages, A.~Pasztor, Ferenc Pittler, J.~Redondo,
  Andreas Ringwald, and K.~Szabo.
\newblock Calculation of the axion mass based on high-temperature lattice
  quantum chromodynamics.
\newblock {\em Nature}, 539:69--71, 11 2016.

\bibitem{Boutan:2017oxg}
Christian~Robert Boutan.
\newblock {\em {A Piezoelectrically Tuned RF-Cavity Search for Dark Matter
  Axions}}.
\newblock PhD thesis, U. Washington, Seattle (main), 2017.

\bibitem{Brady2022}
Anthony~J. Brady, Christina Gao, Roni Harnik, Zhen Liu, Zheshen Zhang, and
  Quntao Zhuang.
\newblock Entangled sensor-networks for dark-matter searches.
\newblock {\em PRX Quantum}, 3:030333, Sep 2022.

\bibitem{Braine2020}
T.~Braine, R.~Cervantes, N.~Crisosto, N.~Du, S.~Kimes, L.~J. Rosenberg,
  G.~Rybka, J.~Yang, D.~Bowring, A.~S. Chou, R.~Khatiwada, A.~Sonnenschein,
  W.~Wester, G.~Carosi, N.~Woollett, L.~D. Duffy, R.~Bradley, C.~Boutan,
  M.~Jones, B.~H. LaRoque, N.~S. Oblath, M.~S. Taubman, J.~Clarke, A.~Dove,
  A.~Eddins, S.~R. O'Kelley, S.~Nawaz, I.~Siddiqi, N.~Stevenson, A.~Agrawal,
  A.~V. Dixit, J.~R. Gleason, S.~Jois, P.~Sikivie, J.~A. Solomon, N.~S.
  Sullivan, D.~B. Tanner, E.~Lentz, E.~J. Daw, J.~H. Buckley, P.~M. Harrington,
  E.~A. Henriksen, and K.~W. Murch.
\newblock Extended search for the invisible axion with the axion dark matter
  experiment.
\newblock {\em Phys. Rev. Lett.}, 124:101303, Mar 2020.

\bibitem{Brouwer_2022}
L.~Brouwer, S.~Chaudhuri, H.-M. Cho, J.~Corbin, W.~Craddock, C.~S. Dawson,
  A.~Droster, J.~W. Foster, J.~T. Fry, P.~W. Graham, R.~Henning, K.~D. Irwin,
  F.~Kadribasic, Y.~Kahn, A.~Keller, R.~Kolevatov, S.~Kuenstner, A.~F. Leder,
  D.~Li, J.~L. Ouellet, K.~M.~W. Pappas, A.~Phipps, N.~M. Rapidis, B.~R. Safdi,
  C.~P. Salemi, M.~Simanovskaia, J.~Singh, E.~C. van Assendelft, K.~van Bibber,
  K.~Wells, L.~Winslow, W.~J. Wisniewski, and B.~A. Young.
\newblock Projected sensitivity of ${\text{dmradio-m}}^{3}$: A search for the
  qcd axion below $1\text{ }\text{ }\mathrm{\ensuremath{\mu}}\mathrm{eV}$.
\newblock {\em Phys. Rev. D}, 106:103008, Nov 2022.

\bibitem{Brubaker2017}
B.~M. Brubaker, L.~Zhong, S.~K. Lamoreaux, K.~W. Lehnert, and K.~A. van Bibber.
\newblock Haystac axion search analysis procedure.
\newblock {\em Phys. Rev. D}, 96:123008, Dec 2017.

\bibitem{carmichael2010statistical}
H.J. Carmichael.
\newblock {\em Statistical Methods in Quantum Optics 2: Non-Classical Fields}.
\newblock Theoretical and Mathematical Physics. Springer Berlin Heidelberg,
  2010.

\bibitem{carmichael2013statistical}
H.J. Carmichael.
\newblock {\em Statistical Methods in Quantum Optics 1: Master Equations and
  Fokker-Planck Equations}.
\newblock Theoretical and Mathematical Physics. Springer Berlin Heidelberg,
  2013.

\bibitem{Caves2012}
Carlton~M. Caves, Joshua Combes, Zhang Jiang, and Shashank Pandey.
\newblock Quantum limits on phase-preserving linear amplifiers.
\newblock {\em Phys. Rev. A}, 86:063802, Dec 2012.

\bibitem{Chakrabarty2017}
S.~S. {Chakrabarty} and P.~{Sikivie}.
\newblock {Effects of an expanding caustic ring on a distribution of stars}.
\newblock In {\em APS April Meeting Abstracts}, January 2017.

\bibitem{Cirac2021}
J.~Ignacio Cirac, David P\'erez-Garc\'{\i}a, Norbert Schuch, and Frank
  Verstraete.
\newblock Matrix product states and projected entangled pair states: Concepts,
  symmetries, theorems.
\newblock {\em Rev. Mod. Phys.}, 93:045003, Dec 2021.

\bibitem{Bai2024}
HAYSTAC Collaboration, Xiran Bai, M.~J. Jewell, J.~Echevers, K.~van Bibber,
  A.~Droster, Maryam~H. Esmat, Sumita Ghosh, Eleanor Graham, H.~Jackson, Claire
  Laffan, S.~K. Lamoreaux, A.~F. Leder, K.~W. Lehnert, S.~M. Lewis, R.~H.
  Maruyama, R.~D. Nath, N.~M. Rapidis, E.~P. Ruddy, M.~Silva-Feaver,
  M.~Simanovskaia, Sukhman Singh, D.~H. Speller, Sabrina Zacarias, and Yuqi
  Zhu.
\newblock Dark matter axion search with haystac phase ii, 2024.

\bibitem{Courant1953}
R.~{Courant} and D.~{Hilbert}.
\newblock {\em {Methods of mathematical physics - Vol.1; Vol.2}}.
\newblock New York: Interscience Publication, 1953, 1953.

\bibitem{Davidson2015}
S.~{Davidson}.
\newblock {Axions: Bose Einstein condensate or classical field?}
\newblock {\em Astroparticle Physics}, 65:101--107, May 2015.

\bibitem{Dine2017}
Michael Dine, Patrick Draper, Laurel Stephenson-Haskins, and Di~Xu.
\newblock Axions, instantons, and the lattice.
\newblock {\em Phys. Rev. D}, 96:095001, Nov 2017.

\bibitem{DINE1983}
Michael {Dine} and Willy {Fischler}.
\newblock {The not-so-harmless axion}.
\newblock {\em Physics Letters B}, 120(1-3):137--141, January 1983.

\bibitem{DINE1981}
Michael Dine, Willy Fischler, and Mark Srednicki.
\newblock A simple solution to the strong cp problem with a harmless axion.
\newblock {\em Physics Letters B}, 104(3):199 -- 202, 1981.

\bibitem{Du:2018uak}
N.~Du et~al.
\newblock {Search for Invisible Axion Dark Matter with the Axion Dark Matter
  Experiment}.
\newblock {\em Phys. Rev. Lett.}, 120(15):151301, 2018.

\bibitem{Duffy2008}
L.~D. {Duffy} and P.~{Sikivie}.
\newblock {Caustic ring model of the MilkyWay halo}.
\newblock {\em \prd}, 78(6):063508, September 2008.

\bibitem{Eberhardt2022}
Andrew Eberhardt, Alvaro Zamora, Michael Kopp, and Tom Abel.
\newblock Single classical field description of interacting scalar fields.
\newblock {\em Phys. Rev. D}, 105:036012, Feb 2022.

\bibitem{Eggemeier_2019}
Benedikt Eggemeier and Jens~C. Niemeyer.
\newblock Formation and mass growth of axion stars in axion miniclusters.
\newblock {\em Physical Review D}, 100(6), Sep 2019.

\bibitem{Eggemeier_2020}
Benedikt Eggemeier, Javier Redondo, Klaus Dolag, Jens~C. Niemeyer, and
  Alejandro Vaquero.
\newblock First simulations of axion minicluster halos.
\newblock {\em Physical Review Letters}, 125(4), Jul 2020.

\bibitem{Erken2012}
O.~{Erken}, P.~{Sikivie}, H.~{Tam}, and Q.~{Yang}.
\newblock {Cosmic axion thermalization}.
\newblock {\em \prd}, 85(6):063520, March 2012.

\bibitem{Fei_2025}
Qinyue Fei, John~D. Silverman, Seiji Fujimoto, Ran Wang, Luis~C. Ho, Manuela
  Bischetti, Stefano Carniani, Michele Ginolfi, Gareth Jones, Roberto Maiolino,
  Wiphu Rujopakarn, N.~M. Förster~Schreiber, Juan~M. Espejo~Salcedo, and L.~L.
  Lee.
\newblock Assessing the dark matter content of two quasar host galaxies at z
  $\sim$ 6 through gas kinematics.
\newblock {\em The Astrophysical Journal}, 980(1):84, February 2025.

\bibitem{Ferreira_2021}
Elisa G.~M. Ferreira.
\newblock Ultra-light dark matter.
\newblock {\em The Astronomy and Astrophysics Review}, 29(1), September 2021.

\bibitem{Flower2022thesis}
Graeme Flower.
\newblock {\em Improving Axion Dark Matter Detection Experiments with a focus
  on the use of Magnonic Systems}.
\newblock PhD thesis, The University of Western Australia, 2022.

\bibitem{Gilman_2025}
Daniel Gilman, Jo~Bovy, Neige Frankel, and Andrew Benson.
\newblock Dark galactic subhalos and the gaia snail.
\newblock {\em The Astrophysical Journal}, 980(1):24, feb 2025.

\bibitem{Gilman2023}
Daniel Gilman, Yi-Ming Zhong, and Jo~Bovy.
\newblock Constraining resonant dark matter self-interactions with strong
  gravitational lenses.
\newblock {\em Phys. Rev. D}, 107:103008, May 2023.

\bibitem{Goodman2025}
C.~Goodman, M.~Guzzetti, C.~Hanretty, L.~J. Rosenberg, G.~Rybka, J.~Sinnis,
  D.~Zhang, John Clarke, I.~Siddiqi, A.~S. Chou, M.~Hollister, S.~Knirck,
  A.~Sonnenschein, T.~J. Caligiure, J.~R. Gleason, A.~T. Hipp, P.~Sikivie,
  M.~E. Solano, N.~S. Sullivan, D.~B. Tanner, R.~Khatiwada, G.~Carosi,
  C.~Cisneros, N.~Du, N.~Robertson, N.~Woollett, L.~D. Duffy, C.~Boutan,
  T.~Braine, E.~Lentz, N.~S. Oblath, M.~S. Taubman, E.~J. Daw, C.~Mostyn, M.~G.
  Perry, C.~Bartram, T.~A. Dyson, S.~Ruppert, M.~O. Withers, C.~L. Kuo, B.~T.
  McAllister, J.~H. Buckley, C.~Gaikwad, J.~Hoffman, K.~Murch, M.~Goryachev,
  E.~Hartman, A.~Quiskamp, and M.~E. Tobar.
\newblock Admx axion dark matter bounds around $3.3\text{ }\text{
  }\mathrm{\ensuremath{\mu}}\mathrm{eV}$ with
  dine-fischler-srednicki-zhitnitsky discovery ability.
\newblock {\em Phys. Rev. Lett.}, 134:111002, Mar 2025.

\bibitem{Guth2015}
A.~H. {Guth}, M.~P. {Hertzberg}, and C.~{Prescod-Weinstein}.
\newblock {Do dark matter axions form a condensate with long-range
  correlation?}
\newblock {\em \prd}, 92(10):103513, November 2015.

\bibitem{hipp2024searchnonvirializedaxions3342}
A.~T. Hipp, A.~Quiskamp, T.~J. Caligiure, J.~R. Gleason, Y.~Han, S.~Jois,
  P.~Sikivie, M.~E. Solano, N.~S. Sullivan, D.~B. Tanner, M.~Goryachev,
  E.~Hartman, M.~E. Tobar, B.~T. McAllister, L.~D. Duffy, T.~Braine, E.~Burns,
  R.~Cervantes, N.~Crisosto, C.~Goodman, M.~Guzzetti, C.~Hanretty, S.~Lee,
  H.~Korandla, G.~Leum, P.~Mohapatra, T.~Nitta, L.~J Rosenberg, G.~Rybka,
  J.~Sinnis, D.~Zhang, C.~Bartram, T.~A. Dyson, C.~L. Kuo, S.~Ruppert, M.~O.
  Withers, M.~H. Awida, D.~Bowring, A.~S. Chou, M.~Hollister, S.~Knirck,
  A.~Sonnenschein, W.~Wester, J.~Brodsky, G.~Carosi, N.~Du, N.~Roberston,
  N.~Woollett, C.~Boutan, A.~M. Jones, B.~H. LaRoque, E.~Lentz, N.~E. Man,
  N.~S. Oblath, M.~S. Taubman, J.~Yang, R.~Khatiwada, John Clarke, I.~Siddiqi,
  A.~Agrawal, A.~V. Dixit, E.~J. Daw, M.~G. Perry, J.~H. Buckley, C.~Gaikwad,
  J.~Hoffman, K.~W. Murch, and J.~Russell.
\newblock Search for non-virialized axions with 3.3-4.2 $\mu$ev mass at
  selected resolving powers, 2024.

\bibitem{Hoskins2016}
J.~Hoskins, N.~Crisosto, J.~Gleason, P.~Sikivie, I.~Stern, N.~S. Sullivan,
  D.~B. Tanner, C.~Boutan, M.~Hotz, R.~Khatiwada, D.~Lyapustin, A.~Malagon,
  R.~Ottens, L.~J. Rosenberg, G.~Rybka, J.~Sloan, A.~Wagner, D.~Will,
  G.~Carosi, D.~Carter, L.~D. Duffy, R.~Bradley, J.~Clarke, S.~O'Kelley, K.~van
  Bibber, and E.~J. Daw.
\newblock Modulation sensitive search for nonvirialized dark-matter axions.
\newblock {\em Phys. Rev. D}, 94:082001, Oct 2016.

\bibitem{jaeckel2022reporttopicalgroupwave}
Joerg Jaeckel, Gray Rybka, and Lindley Winslow.
\newblock Report of the topical group on wave dark matter for snowmass 2021,
  2022.

\bibitem{Jewell2023}
M.~J. Jewell, A.~F. Leder, K.~M. Backes, Xiran Bai, K.~van Bibber, B.~M.
  Brubaker, S.~B. Cahn, A.~Droster, Maryam~H. Esmat, Sumita Ghosh, Eleanor
  Graham, Gene~C. Hilton, H.~Jackson, Claire Laffan, S.~K. Lamoreaux, K.~W.
  Lehnert, S.~M. Lewis, M.~Malnou, R.~H. Maruyama, D.~A. Palken, N.~M. Rapidis,
  E.~P. Ruddy, M.~Simanovskaia, Sukhman Singh, D.~H. Speller, Leila~R. Vale,
  H.~Wang, and Yuqi Zhu.
\newblock New results from haystac's phase ii operation with a squeezed state
  receiver.
\newblock {\em Phys. Rev. D}, 107:072007, Apr 2023.

\bibitem{Kim1979}
Jihn~E. Kim.
\newblock Weak-interaction singlet and strong cp invariance.
\newblock {\em Phys. Rev. Lett.}, 43:103--107, Jul 1979.

\bibitem{BREAD2024}
Stefan Knirck, Gabe Hoshino, Mohamed~H. Awida, Gustavo~I. Cancelo, Martin
  Di~Federico, Benjamin Knepper, Alex Lapuente, Mira Littmann, David~W. Miller,
  Donald~V. Mitchell, Derrick Rodriguez, Mark~K. Ruschman, Matthew~A. Sawtell,
  Leandro Stefanazzi, Andrew Sonnenschein, Gary~W. Teafoe, Daniel Bowring,
  G.~Carosi, Aaron Chou, Clarence~L. Chang, Kristin Dona, Rakshya Khatiwada,
  Noah~A. Kurinsky, Jesse Liu, Cristi\'an Pena, Chiara~P. Salemi, Christina~W.
  Wang, and Jialin Yu.
\newblock First results from a broadband search for dark photon dark matter in
  the 44 to $52\text{ }\text{ }\mathrm{\ensuremath{\mu}}\mathrm{eV}$ range with
  a coaxial dish antenna.
\newblock {\em Phys. Rev. Lett.}, 132:131004, Mar 2024.

\bibitem{Kopp2022}
Michael Kopp, Vasileios Fragkos, and Igor Pikovski.
\newblock Nonclassicality of axionlike dark matter through gravitational
  self-interactions.
\newblock {\em Phys. Rev. D}, 106:043517, Aug 2022.

\bibitem{Lentz2020c}
Erik~W. {Lentz}, Leon {Lettermann}, Thomas~R. {Quinn}, and Leslie~J
  {Rosenberg}.
\newblock {Mixed State Dynamics with Non-Local Interactions}.
\newblock {\em arXiv e-prints}, page arXiv:2002.09173, February 2020.

\bibitem{Lentz2019}
Erik~W. {Lentz}, Thomas~R. {Quinn}, and Leslie~J. {Rosenberg}.
\newblock {Axion structure formation - I: the co-motion picture}.
\newblock {\em Monthly Notices of the Royal Astronomical Society},
  485(2):1809--1821, May 2019.

\bibitem{Lentz2020}
Erik~W Lentz, Thomas~R Quinn, and Leslie~J Rosenberg.
\newblock {Axion structure formation – II. The wrath of collapse}.
\newblock {\em Monthly Notices of the Royal Astronomical Society},
  493(4):5944--5971, 03 2020.

\bibitem{Lentz2020b}
Erik~W. Lentz, Thomas~R. Quinn, and Leslie~J. Rosenberg.
\newblock Condensate dynamics with non-local interactions.
\newblock {\em Nuclear Physics B}, page 114937, 2020.

\bibitem{Levkov_2018}
D.G. Levkov, A.G. Panin, and I.I. Tkachev.
\newblock Gravitational bose-einstein condensation in the kinetic regime.
\newblock {\em Physical Review Letters}, 121(15), Oct 2018.

\bibitem{Malnou2019}
M.~Malnou, D.~A. Palken, B.~M. Brubaker, Leila~R. Vale, Gene~C. Hilton, and
  K.~W. Lehnert.
\newblock Squeezed vacuum used to accelerate the search for a weak classical
  signal.
\newblock {\em Phys. Rev. X}, 9:021023, May 2019.

\bibitem{Marsh:2015xka}
David J.~E. Marsh.
\newblock {Axion Cosmology}.
\newblock {\em Phys. Rept.}, 643:1--79, 2016.

\bibitem{Meneghetti2020}
Massimo {Meneghetti}, Guido {Davoli}, Pietro {Bergamini}, Piero {Rosati},
  Priyamvada {Natarajan}, Carlo {Giocoli}, Gabriel~B. {Caminha}, R.~Benton
  {Metcalf}, Elena {Rasia}, Stefano {Borgani}, Francesco {Calura}, Claudio
  {Grillo}, Amata {Mercurio}, and Eros {Vanzella}.
\newblock {An excess of small-scale gravitational lenses observed in galaxy
  clusters}.
\newblock {\em Science}, 369(6509):1347--1351, September 2020.

\bibitem{Minor2021}
Quinn Minor, Sophia Gad-Nasr, Manoj Kaplinghat, and Simona Vegetti.
\newblock An unexpected high concentration for the dark substructure in the
  gravitational lens sdssj0946+1006.
\newblock {\em Monthly Notices of the Royal Astronomical Society},
  507(2):1662--1683, 08 2021.

\bibitem{MTW}
C.~W. {Misner}, K.~S. {Thorne}, and J.~A. {Wheeler}.
\newblock {\em {Gravitation}}.
\newblock 1973.

\bibitem{Mocz2018}
Philip Mocz, Lachlan Lancaster, Anastasia Fialkov, Fernando Becerra, and
  Pierre-Henri Chavanis.
\newblock Schr\"odinger-poisson--vlasov-poisson correspondence.
\newblock {\em Phys. Rev. D}, 97:083519, Apr 2018.

\bibitem{Neumann1927}
J.~von Neumann.
\newblock Wahrscheinlichkeitstheoretischer aufbau der quantenmechanik.
\newblock {\em Nachrichten von der Gesellschaft der Wissenschaften zu
  Göttingen, Mathematisch-Physikalische Klasse}, 1927:245--272, 1927.

\bibitem{Niemeyer_2020}
Jens~C. Niemeyer.
\newblock Small-scale structure of fuzzy and axion-like dark matter.
\newblock {\em Progress in Particle and Nuclear Physics}, 113:103787, July
  2020.

\bibitem{Oman2015}
Kyle~A. Oman, Julio~F. Navarro, Azadeh Fattahi, Carlos~S. Frenk, Till Sawala,
  Simon D.~M. White, Richard Bower, Robert~A. Crain, Michelle Furlong, Matthieu
  Schaller, Joop Schaye, and Tom Theuns.
\newblock The unexpected diversity of dwarf galaxy rotation curves.
\newblock {\em Monthly Notices of the Royal Astronomical Society},
  452(4):3650--3665, 08 2015.

\bibitem{ORUS2014}
Román Orús.
\newblock A practical introduction to tensor networks: Matrix product states
  and projected entangled pair states.
\newblock {\em Annals of Physics}, 349:117--158, 2014.

\bibitem{PQ1977}
R.~D. Peccei and Helen~R. Quinn.
\newblock Cp conservation in the presence of pseudoparticles.
\newblock {\em Phys. Rev. Lett.}, 38:1440--1443, 1977.

\bibitem{PRESKILL1983}
John Preskill, Mark~B. Wise, and Frank Wilczek.
\newblock Cosmology of the invisible axion.
\newblock {\em Physics Letters B}, 120(1):127 -- 132, 1983.

\bibitem{roberts2024}
M.~Grant Roberts, Manoj Kaplinghat, Mauro Valli, and Hai-Bo Yu.
\newblock Gravothermal collapse and the diversity of galactic rotation curves,
  2024.

\bibitem{Schive_2014}
Hsi-Yu Schive, Tzihong Chiueh, and Tom Broadhurst.
\newblock Cosmic structure as the quantum interference of a coherent dark wave.
\newblock {\em Nature Physics}, 10(7):496–499, Jun 2014.

\bibitem{SHIFMAN1980}
M.A. Shifman, A.I. Vainshtein, and V.I. Zakharov.
\newblock Can confinement ensure natural cp invariance of strong interactions?
\newblock {\em Nuclear Physics B}, 166(3):493 -- 506, 1980.

\bibitem{Sikivie1983}
P.~Sikivie.
\newblock Experimental tests of the "invisible" axion.
\newblock {\em Phys. Rev. Lett.}, 51:1415--1417, Oct 1983.

\bibitem{Sikivie1999a}
P.~{Sikivie}.
\newblock {Caustic ring singularity}.
\newblock {\em \prd}, 60(6):063501, September 1999.

\bibitem{Sikivie1999b}
P.~{Sikivie}.
\newblock {Velocity peaks and caustic rings}.
\newblock {\em Nuclear Physics B Proceedings Supplements}, 72:110--113, March
  1999.

\bibitem{Sikivie2011}
P.~{Sikivie}.
\newblock {The emerging case for axion dark matter}.
\newblock {\em Physics Letters B}, 695:22--25, January 2011.

\bibitem{Sikivie2009}
P.~{Sikivie} and Q.~{Yang}.
\newblock {Bose-Einstein Condensation of Dark Matter Axions}.
\newblock {\em Physical Review Letters}, 103(11):111301, September 2009.

\bibitem{Turner1990}
Michael~S. Turner.
\newblock Periodic signatures for the detection of cosmic axions.
\newblock {\em Phys. Rev. D}, 42:3572--3575, Nov 1990.

\bibitem{Veltmaat_2018}
Jan Veltmaat, Jens~C. Niemeyer, and Bodo Schwabe.
\newblock Formation and structure of ultralight bosonic dark matter halos.
\newblock {\em Physical Review D}, 98(4), Aug 2018.

\bibitem{Veltmaat_2020}
Jan Veltmaat, Bodo Schwabe, and Jens~C. Niemeyer.
\newblock Baryon-driven growth of solitonic cores in fuzzy dark matter halos.
\newblock {\em Physical Review D}, 101(8), Apr 2020.

\bibitem{Verstraete2004}
F.~{Verstraete} and J.~I. {Cirac}.
\newblock {Renormalization algorithms for Quantum-Many Body Systems in two and
  higher dimensions}.
\newblock {\em arXiv e-prints}, pages cond--mat/0407066, July 2004.

\bibitem{Vool2017}
Uri Vool and Michel Devoret.
\newblock Introduction to quantum electromagnetic circuits.
\newblock {\em International Journal of Circuit Theory and Applications},
  45(7):897--934, 2017.

\bibitem{Wald1984}
R.~M. {Wald}.
\newblock {\em {General relativity}}.
\newblock 1984.

\bibitem{Weinberg1978}
Steven Weinberg.
\newblock A new light boson?
\newblock {\em Phys. Rev. Lett.}, 40:223--226, Jan 1978.

\bibitem{Wilczek1978}
F.~Wilczek.
\newblock Problem of strong $p$ and $t$ invariance in the presence of
  instantons.
\newblock {\em Phys. Rev. Lett.}, 40:279--282, Jan 1978.

\bibitem{Xiao2024}
Mengyuan {Xiao}, Pascal~A. {Oesch}, David {Elbaz}, Longji {Bing}, Erica~J.
  {Nelson}, Andrea {Weibel}, Garth~D. {Illingworth}, Pieter {van Dokkum},
  Rohan~P. {Naidu}, Emanuele {Daddi}, Rychard~J. {Bouwens}, Jorryt {Matthee},
  Stijn {Wuyts}, John {Chisholm}, Gabriel {Brammer}, Mark {Dickinson}, Benjamin
  {Magnelli}, Lucas {Leroy}, Daniel {Schaerer}, Thomas {Herard-Demanche},
  Seunghwan {Lim}, Laia {Barrufet}, Ryan {Endsley}, Yoshinobu {Fudamoto},
  Carlos {G{\'o}mez-Guijarro}, Rashmi {Gottumukkala}, Ivo {Labb{\'e}}, Dan
  {Magee}, Danilo {Marchesini}, Michael {Maseda}, Yuxiang {Qin}, Naveen~A.
  {Reddy}, Alice {Shapley}, Irene {Shivaei}, Marko {Shuntov}, Mauro {Stefanon},
  Katherine~E. {Whitaker}, and J.~Stuart~B. {Wyithe}.
\newblock {Accelerated formation of ultra-massive galaxies in the first billion
  years}.
\newblock {\em \nat}, 635(8038):311--315, November 2024.

\bibitem{Zhang_2025}
Xingyu Zhang, Hai-Bo Yu, Daneng Yang, and Ethan~O. Nadler.
\newblock The gd-1 stellar stream perturber as a core-collapsed
  self-interacting dark matter halo.
\newblock {\em The Astrophysical Journal Letters}, 978(2):L23, jan 2025.

\bibitem{Zhitnitsky1980}
A.~R. Zhitnitsky.
\newblock {On Possible Suppression of the Axion Hadron Interactions. (In
  Russian)}.
\newblock {\em Sov. J. Nucl. Phys.}, 31:260, 1980.
\newblock [Yad. Fiz.31,497(1980)].

\bibitem{Zhong2018}
L.~{Zhong}, S.~{Al Kenany}, K.~M. {Backes}, B.~M. {Brubaker}, S.~B. {Cahn},
  G.~{Carosi}, Y.~V. {Gurevich}, W.~F. {Kindel}, S.~K. {Lamoreaux}, K.~W.
  {Lehnert}, S.~M. {Lewis}, M.~{Malnou}, R.~H. {Maruyama}, D.~A. {Palken},
  N.~M. {Rapidis}, J.~R. {Root}, M.~{Simanovskaia}, T.~M. {Shokair}, D.~H.
  {Speller}, I.~{Urdinaran}, and K.~A. {van Bibber}.
\newblock {Results from phase 1 of the HAYSTAC microwave cavity axion
  experiment}.
\newblock {\em \prd}, 97(9):092001, May 2018.

\end{thebibliography}

\end{document}